\documentclass[prd,showpacs,twocolumn,aps,superscriptaddress,nofootinbib]{revtex4-2}
\usepackage{amssymb,amsmath,bm}
\usepackage{graphicx}
\usepackage[colorlinks=true, pdfstartview=FitV, linkcolor=blue, citecolor=blue, urlcolor=blue]{hyperref}

\usepackage{multirow}

\usepackage{graphicx} 

\newcommand{\bb}{{\boldsymbol b}}
\newcommand{\br}{{\boldsymbol r}}
\newcommand{\bx}{{\boldsymbol x}}

\newcommand{\bk}{{\boldsymbol k}}
\newcommand{\bv}{{\boldsymbol v}}
\newcommand{\bu}{{\boldsymbol u}}
\newcommand{\bB}{{\boldsymbol B}}
\newcommand{\bE}{{\boldsymbol E}}
\newcommand{\be}{{\boldsymbol e}}

\newcommand{\bj}{{\boldsymbol j}}
\newcommand{\bnabla}{{\boldsymbol \nabla}}

\newcommand{\bO}{{\boldsymbol \Omega}}
\newcommand{\bo}{{\boldsymbol \omega}}

\renewcommand\a{\alpha}

\renewcommand\d{\delta}
\renewcommand\k{\kappa}
\renewcommand\l{\lambda}
\renewcommand\r{\rho}
\renewcommand\t{\tau}

\renewcommand\c{\chi}

\renewcommand\o{\omega}
\newcommand\e{\epsilon}
\newcommand\g{\gamma}
\newcommand\z{\zeta}
\newcommand\m{\mu}
\newcommand\n{\nu}
\newcommand\x{\xi}
\newcommand\p{\pi}
\newcommand\h{\theta}
\newcommand\s{\sigma}

\newcommand\w{\eta}
\newcommand\ve{\varepsilon}

\renewcommand\O{\Omega}

\newcommand\D{\Delta}
\newcommand\G{\Gamma}

\newcommand\pt{\partial}

\newcommand\ls{\left[}
\newcommand\rs{\right]}

\newcommand{\lan}{\langle}
\newcommand{\ran}{\rangle}

\newcommand{\eq}[1]{Eq.~(\ref{#1})}
\newcommand{\eqs}[2]{Eqs.~(\ref{#1})-(\ref{#2})}

\newcommand\pll{\parallel}

\begin{document}

\title{Rotation catalyzed chiral magnetovortical instability}
\author{Shuai Wang}
\affiliation{Physics Department and Center for Particle Physics and Field Theory, Fudan University, Shanghai 200438, China}
\author{Xu-Guang Huang}
\email{huangxuguang@fudan.edu.cn}
\affiliation{Physics Department and Center for Particle Physics and Field Theory, Fudan University, Shanghai 200438, China}
\affiliation{Key Laboratory of Nuclear Physics and Ion-beam Application (MOE), Fudan University, Shanghai 200433, China}
\affiliation{Shanghai Research Center for Theoretical Nuclear Physics, NSFC and Fudan University, Shanghai 200438, China}

\begin{abstract}
We demonstrate that a background rotation significantly catalyzes the chiral magnetovortical instability in chiral magnetohydrodynamics. The rotation splits the linearly polarized Alfv\'en wave into two circularly polarized magneto-Coriolis waves, one of which exhibits a lower frequency than the original Alfv\'en wave. We find that this low-frequency magneto-Coriolis wave is always unstable in the presence of  even a weak chiral vortical effect. This instability may enable new dynamo mechanism applicable to various rotating chiral plasmas.
\end{abstract}

\maketitle

\section{Introduction}\label{intro}
Chiral matter exhibits a variety of intriguing anomalous transport phenomena that originate from the chiral anomaly and are absent in normal matter. Prominent examples of such phenomena include the chiral magnetic effect (CME)~\cite{PhysRevD.22.3080, KHARZEEV2008227, PhysRevD.78.074033} and the chiral vortical effect (CVE)~\cite{PhysRevD.20.1807, JohannaErdmenger_2009,Banerjee:2008th,Son:2009tf}, which are electric currents along the magnetic and vortical fields, respectively. These novel effects have been discussed in a broad range of contexts, including the quark-gluon plasma (QGP) in heavy-ion collisions, the early Universe, supernovae and neutron stars, and Weyl/Dirac semimetals; see Refs.~\cite{Huang:2015oca,Liu:2020ymh,Kharzeev:2020jxw,Chernodub:2021nff} for reviews. 

From a theoretical perspective, the low-energy behavior of chiral matter can be described by chiral (anomalous) hydrodynamics or chiral magnetohydrodynamics (MHD)~\cite{Hattori:2022hyo}, in which the anomalous electric currents from the CME and CVE are encoded. A number of novel wave modes have been identified in chiral MHD, such as the chiral magnetic wave~\cite{PhysRevD.83.085007}, chiral vortical wave~\cite{PhysRevD.92.071501}, chiral electric wave~\cite{PhysRevLett.110.232302}, chiral Alfv\'en wave~\cite{PhysRevLett.115.141601}, chiral heat wave~\cite{Chernodub:2015gxa}, and chiral shock wave~\cite{PhysRevLett.118.181601}. Moreover, some of these wave modes can become unstable under certain conditions. For instance, the CME can induce an instability known as the chiral plasma instability or chiral dynamo instability~\cite{PhysRevLett.79.1193, PhysRevLett.111.052002}, while the CVE can trigger an MHD instability called the chiral magnetovortical instability (CMVI)~\cite{PhysRevD.109.L121302}, which requires consideration of the mutual evolution of the magnetic and vortical fields. Additionally, pure CVE-induced instability has also been recently discussed~\cite{Wang:2025bfe}. Furthermore, these instabilities may provide new dynamo mechanisms that can operate in various physical environments, such as Weyl semimetals~\cite{PhysRevLett.121.176603,Qiu:2016hzd,Amitani:2022pmu,Nishida:2023}, the early Universe~\cite{PhysRevLett.79.1193,Giovannini:1997eg,Boyarsky:2011uy,Boyarsky:2012ex,Tashiro:2012mf,Dvornikov:2016jth,Gorbar:2016klv,Brandenburg:2017rcb,Schober:2017cdw},  supernovae and neutron stars~\cite{PhysRevLett.111.052002,Schober:2017cdw,Ohnishi:2014uea,Yamamoto:2015gzz,Grabowska:2014efa,Dvornikov:2014uza,Sigl:2015xva,Matsumoto:2022lyb,Schober:2021yav,Brandenburg:2023aco,Dehman:2024zqc,Dehman:2025jsv}, and heavy-ion collisions~\cite{Akamatsu:2015kau,Hirono:2015rla,Xia:2016any,Tuchin:2017vwb,Schlichting:2022fjc}.

As mentioned above, the CVE plays a vital role in inducing novel instabilities, highlighting the importance of the vortical field. The simplest vortical field is a global rotation, which has been extensively explored in conventional hydrodynamics~\cite{Landau, Davidson_2016}. This is primarily because rotation is an extremely common feature in many systems. Typically, it is convenient to consider a rotating fluid in a co-rotating frame (a non-inertial frame), where the Coriolis and centrifugal forces emerge. For example, in a rotating fluid, the Coriolis force acts as a restoring force, supporting the so-called inertial waves~\cite{Landau}. These waves propagate with a direction-dependent frequency and are damped by viscous effects. In the context of MHD, a well-known wave is the Alfv\'en wave~\cite{Davidson_2016}, which is excited by magnetic tension and is damped by viscous and resistive effects. When considering rotating MHD, inertial waves couple with Alfv\'en waves to form magneto-Coriolis (MC) waves~\cite{1954ApJ...119..647L,1955ApJ...121..481L,DJAcheson_1973, PhysRevLett.104.074501, Rax_Gueroult_Fisch_2023}. These coupled waves can be understood as the splitting of linearly polarized Alfv\'en waves into circularly polarized waves by the rotation~\cite{PhysRevLett.104.074501, 1954ApJ...119..647L, 1955ApJ...121..481L, Davidson_2013}, which play an important role in astrophysics.

Regarding instabilities in rotating matter, a renowned example is the magnetorotational instability (MRI), which plays an essential role in understanding the transport of angular momentum in astronomical accretion disks~\cite{RevModPhys.70.1, PhysRevLett.129.115001}. Although in this case the fluid is subject to differential rotation rather than rigid rotation, it is clear that both the magnetic field and the rotation (whether rigid or differential) play crucial roles in shaping the evolution of MHD systems. Recently, a rotational instability of the magnetic field has been discussed~\cite{Das:2025kgq}.

In this paper, we investigate the impact of global rotation on the CMVI. We demonstrate that rotation results in a hybridization of chiral Alfvén waves with conventional inertial and Alfvén waves, thereby inducing a novel type of CMVI. Unlike the CMVI identified in Ref.~\cite{PhysRevD.109.L121302} for non-rotating plasmas, where the onset condition is given by $\xi_\omega > \sqrt{\r}$ (with $\xi_\omega$ being the CVE coefficient and $\r$ the mass density), in the presence of global rotation, the CMVI can be significantly catalyzed by the Coriolis force. We adhere to the notations used in Ref.~\cite{PhysRevD.109.L121302}.

\section{MHD in a rotating frame}\label{sec:mhdro}
We begin by reviewing the rotational effects in conventional MHD. We consider a conducting fluid rotating with constant angular velocity $\bO$. We work in the co-rotating reference frame (a non-inertial reference frame). The governing non-relativistic MHD equations are given by \cite{DJAcheson_1973, Davidson_2016} (see Appendix \ref{derivofeqns} for a derivation)
\begin{align}
\label{NS-rot-inEMF}
\pt_{t}\bv+(\bv\cdot \bnabla)\bv & +2\bm{\O} \times \bv  = -\bm{\nabla} P  + \frac{1}{\r} (\bB\cdot\bnabla)\bB,   \\
 \label{pt-B}
 &\pt_{t}\bB  = \bnabla \times (\bv\times\bB) 
               +\eta \bnabla^2\bB, 
\end{align}
where $\r$ is the mass density, $\bv(t,\br)$ is the velocity field ($\br$ is the position vector from the rotating axis), $\bB(t,\br)$ is the magnetic field, $P =p/\r-|\bO\times\br|^2/2+\bB^2/2\r$ is the generalized normalized pressure with $p$ the thermodynamic pressure, $\eta$ is the electric resistivity, $-2\bO\times\bv$ is the Coriolis force, and $-\bnabla(|\bO\times\br|^2/2)=\bO\times(\bO\times\br)$ is the centrifugal force. We have assumed $\r$ to be a constant which implies the incompressiblity of the fluid, 
\begin{align}
	\label{divergenceless}
\bnabla\cdot\bv=0.
\end{align}
Incompressibility is not crucial for our discussions and the results below can be readily extended to compressible case. Without loss of generality, the background rotating field $\bO$ is set to be along the $z$ direction, such that $\bO=\O \hat {\bm{z}}$ with $\O>0$. In \eq{NS-rot-inEMF} and \eq{pt-B}, all physical quantities are defined in the rotating frame (see Appendix \ref{derivofeqns} for detailed definitions).

Equations (\ref{NS-rot-inEMF}) and (\ref{pt-B}) allow an equilibrium state with constant magnetic field $\bB_0$ and vanishing velocity. This is achieved when the thermodynamic pressure gradient balances the centrifugal force. Let us consider small perturbations around this equilibrium state and examine how the perturbations evolve. We decompose the magnetic field into two parts $\bB=\bB_{0}+\bb$ and the velocity field into $\bv=\bm 0+\bv$, where $\bb,\bv$ are treated as small quantities with $|\bb|/|\bB_0|,|\bv|\sim \d \ll 1$. We assume $\bB_{0}=B_{0} \hat{\bm{z}}$ with $B_{0}>0$ which is aligned with the direction of $\bO$. 
Linearizing \eqs{NS-rot-inEMF}{pt-B} to first order in $\d$, we obtain
\begin{align}
\label{linear-v}
\pt_{t}\bv &=- 2\bm{\O} \times \bv -\bm{\nabla} P 
  +\frac{ (\bB_{0}\cdot\bnabla)\bb  }{\r}, \\
\label{linear-b}
\pt_{t}\bb &=\eta\bnabla^2\bb+ (\bB_{0}\cdot\bnabla)\bv,
\end{align}
where the imcompressibility condition (\ref{divergenceless}) is used.
To determine the eigenmodes, we substitute the plane-wave ansatz for the fluctuations, $\bb,\bv= \bm{f} _{b,v}e^{i(\bk\cdot\bx-\o t)}$ where $\bm{f}_{b,v}$ are amplitude vectors, and apply curl to Eq.~(\ref{linear-v}). This yields 
\begin{align}
\label{linear-v-k}
\o\bk\times\bm{f}_{v}-2i(\bO\cdot\bk)\bm{f}_{v} & =
-\frac{1}{\r}(\bB_{0}\cdot\bk)\bk \times \bm{f}_{b},\\
\label{linear-b-kddd}
(-i\o+\eta\bk^2)\bm{f}_{b} &= i (\bB_{0}\cdot\bk) \bm{f}_{v} ,
\end{align}
by virtue of $\bnabla\cdot\bv=\bnabla\cdot\bb=0$. The dispersion relations are obtained from solving
\begin{align}
&\o^2 +(i\eta\bk^2-\l\frac{2\bO \cdot \bk}{k})\o 
 - i\l\eta\bk^2 \frac{2\bO\cdot \bk}{k}
   -\frac{(\bB_{0}\cdot\bk)^2}{\r} = 0 ,
   \label{dis-rotating} 
\end{align} 
where $\l=\pm 1$ and $k\equiv|\bk|$. For brevity, we define
\begin{align}
\o_{I}\equiv 2\frac{\bO\cdot\bk}{k},\quad \o_{A}\equiv \frac{\bB_{0}\cdot\bk}{\sqrt{\r}}.
\end{align}
They are frequencies of pure inertial wave and Alfvén wave, respectively. The solutions to \eq{dis-rotating} are then given by 
\begin{equation}
\o  =-i\frac{\g_\w } {2} +\l\frac{\o_{I}}{2}  
\pm \sqrt{ (\frac{i\g_\w}{2}+\l\frac{\o_{I}}{2})^2+\o_{A}^2 }   ,
\label{solution-O-B}
\end{equation}
with $\g_\eta\equiv \eta\bk^2$ the magnetic diffusion rate. Equation (\ref{solution-O-B}) reveals that Alfvén waves and inertial waves are coupled in a nonlinear manner, and these coupled waves are damped by electric resistivity $\eta$, which typically prevents the occurrence of instability. 

For ideal MHD, $\g_\w=0$, the frequency (\ref{solution-O-B}) becomes $\o=\pm\o_{f,s}$,  with 
\begin{align}
	\o_{f} &=  \o_{I}/2+(\o_{I}^2/4+\o_{A}^2)^{1/2} ,
	\label{fast-MC}  \\
	\o_{s} &= -\o_{I}/2 + (\o_{I}^2/4+\o_{A}^2)^{1/2}, 
	\label{slow-MC}
\end{align}
indicating that the rotation splits the usal Alfv\'en wave of frequency $\o_A$ into two distinct waves: a fast one with frequency $\o_f$ (for $\o_I>0$) and a slow one with frequency $\o_s$~\cite{1954ApJ...119..647L,1955ApJ...121..481L,DJAcheson_1973,Rax_Gueroult_Fisch_2023}. We refer to these as the magneto-Coriolis (MC) waves, as the restoring force is the combination of Lorentz force (more precisely, magnetic tension induced by Lorentz force) and Coriolis forces. For fast MC wave, the Coriolis force adds to the Lorentz force, leading to a larger frequency $\o_f > \o_A$, while for the slow MC wave, the Coriolis force acts opposite to the Lorentz force, resulting a smaller frequency $\o_s < \o_A$. Note that $\o_f\o_s=\o_A^2$. Unlike the usual Alfv\'en wave at zero rotation, which is linearly polarized, the MC waves are circularly polarized. For weak rotation, where $\o_I\ll \o_A$, we have $\o_f\approx \o_A+\o_I/2$ and $\o_s\approx\o_A-\o_I/2$, showing that the MC waves are predominantly Alfv\'en waves with a small inertial wave component. For strong rotation, where $\o_{I} \gg \o_{A}$, we have $\o_f\approx\o_I$, representing the the inertial wave, and $\o_{s}\approx \o_{A}^2/\o_{I} \ll \o_{A}$, which is very slow and sometimes termed the magnetostrophic wave~\cite{Davidson_2013, Bardsley_Davidson_2016,Sreenivasan_Narasimhan_2017}. 

Finally, let us clarify the physical meaning of $\l=\pm$. Set the helicity basis with $\bm{e}_{3}(\bk)= \hat{\bk}=\bk/k$ (for $k>0$) along the wave vector direction, and the right-hand and left-hand helicity basis vectors $\bm{e}_{\pm}(\bk)$, which satisfy $\hat{\bk}\times \bm{e}_{\pm}(\bk)=\mp i\bm{e}_{\pm}(\bk), \hat{\bk} \cdot \bm{e}_{\pm}(\bk)=0, \bm{e}_{\pm}(\bk)\cdot \bm{e}_{\pm}(\bk)^*=1$, and $\bm{e}_{\pm}(\bk)\cdot\bm{e}_{\mp}(\bk)^*=0$. Due to the conditions $\bnabla\cdot\bv=0=\bnabla\cdot\bb$, $\bm f_v$ and $\bm f_b$ are transverse to $\bm{e}_3$ and thus can be expanded as $\bm{f}_{v,b}=f_{v,b+}\bm{e}_{+}+f_{v,b-}\bm{e}_{-}$. Then \eqs{linear-v-k}{linear-b-kddd} simplify to 
\begin{align}
& \o f_{v\pm} \pm \o_I f_{v\pm}+\o_A f_{b\pm}/\sqrt{\r}=0,\\
& (\o+i\g_\w)f_{b\pm}/\sqrt{\r}+\o_A f_{v\pm}=0.
\end{align}
From this, we obtain the following equation for $\o$,
\begin{align}
&\o^2 +(i\g_\w\pm\o_I)-\o_A^2\pm i\g_\w\o_I=0.
\end{align} 
Comparing to \eq{dis-rotating}, one identifies that $\l=+1$ in the dispersion relation (\ref{solution-O-B}) corresponds to left-hand waves and $\l=-1$ corresponds to right-hand waves.

\section{Chiral MHD in a Rotating Frame}\label{sec:cmhd}
Chiral MHD is dramatically different from the conventional MHD due to the chiral anomalous effects, such as the CME and CVE, in the electric current
\begin{align}
	\label{currentinrf}
 \bj_B = \xi_{B}\bB, \quad \bj_\o = \xi_{\o}(\bo+2\bO),
 \end{align}
where $\xi_{B,\o}\propto\m_5$ are the CME and CVE conductivities with $\m_5$ the chiral chemical potential (without loss of generality, we will henceforth assume $\x_{B,\o}>0$ in all subsequent discussions), and $\bo\equiv\bnabla\times \bv$ is the vorticity. Note that there is an additional $2\bO$ contribution to the CVE ;  see Appendix \ref{derivofeqns} for a derivation. We emphasize again that physical quantities are defined in the rotating frame. In order to simplify the notation, we will frequenctly use the following dimensionless quantities,
\begin{align}
	\label{currentinrf}
 \x'_\o\equiv \frac{\x_\o}{\sqrt{\r}},\, \bB'_0 \equiv \frac{\bB_0}{\sqrt{\r}},\,  \bb' \equiv \frac{\bb}{\sqrt{\r}},\, \bm f'_b \equiv \frac{\bm f_b}{\sqrt{\r}}.
 \end{align}

For the purpose of our analysis, we focus on the non-relativistic limit, as it offers a more transparent understanding of the origin of instabilities. A similar analysis can be performed  for the relativistic case as well. The governing equations are \eq{NS-rot-inEMF} and (see Appendix \ref{derivofeqns} for a derivation)
\begin{equation}
 \label{pt-B-chiral}
 \pt_{t}\bB = \eta \bnabla^2\bB+\bnabla \times (\bv\times\bB)  +\eta \xi_{B}\bnabla\times\bB +\eta\xi_{\o}\bnabla\times\bo.
\end{equation}
Although the form of Eq. (\ref{pt-B-chiral}) is the same as Eq. (3) in Ref.~\cite{PhysRevD.109.L121302}, the quantities in Eq. (\ref{pt-B-chiral}) are defined in the rotating frame. Making a linear analysis for Eq. (\ref{pt-B-chiral}) similar to Sec.~\ref{sec:mhdro}, one obtains
\begin{equation}
\label{lin-b-chiral}
    \pt_{t}\bb= \eta \bnabla^2 \bb + (\bB_{0}\cdot\bnabla)\bv
  + \eta \xi_{B} \bnabla\times\bb-\eta \xi_{\o} \bnabla^2\bv. 
\end{equation}
After substituting the plane wave ansatz into Eq. (\ref{lin-b-chiral}), it transforms into an algebraic form
\begin{align}
\label{lin-b-chiral-k}
 & (\eta\bk^2-i\o)\bm{f}_{b}-i\eta\xi_{B}\bk\times\bm{f}_{b}
  = [i\bB_{0}\cdot\bk + \eta \xi_{\o}\bk^2 ]\bm{f}_{v}. 
\end{align}
Finally, combining Eqs.~(\ref{linear-v-k}) and (\ref{lin-b-chiral-k}), we find that the dispersion relations for the eigenmodes are determined by
\begin{align}
\label{chiral-dispersion-total}
& \o^2+[i\eta k( k +\l \xi_{B})-\l \o_{I} ]\o  \nonumber   \\
&\quad   -i\eta k(\xi_{B} +\l k )\o_{I}   
  -\o_{A}^2+i\eta k^2 \o_{A}\xi_{\o}^{\prime} =0  ,
\end{align}
where $\l=\pm 1$. Before delving into the solutions of Eq.~(\ref{chiral-dispersion-total}), let us makes a few remarks about it. \\
(1) The last term in Eq.~(\ref{chiral-dispersion-total}) corresponds to the contribution from the so-called chiral  Alfv\'en wave~\cite{PhysRevLett.115.141601}. The frequency of a pure chiral Alfv\'en wave is 
\begin{align}
\o_{CA}\equiv\x_\o'\o_A.
\end{align}
This is most easily seen by turning off the CME and considering the large $\eta$ limit. In this case, the solution is $\o=\l\o_I-\x_\o'\o_A$, representing a linear combination of chiral Alfv\'en wave and inertial wave. \\
(2) When $\bO=0$, Eq.~(\ref{chiral-dispersion-total}) reduces to the situation studied in Ref.~\cite{PhysRevD.109.L121302}. In this case, it was known that a chiral magnetovortical instability (CMVI) emerges when $\x_\o'>1$~\cite{PhysRevD.109.L121302}. To understand this, we turn off the CME and rotation and rewrite  \eq{chiral-dispersion-total} as 
\begin{align}
\label{chiralddd-dispersion-total}
(\o-\o_A)(\o+\o_A)+i\g_\eta (\o+\o_{CA})=0.
\end{align}
The sign of the last term on the left-hand side determines whether instability can arise. When $|\o_{CA}|\leq|\o_A|$, the solutions correspond to damped Alfv\'en waves modified by the CVE. Unstable modes appear only when $|\o_{CA}|>|\o_A|$, which are the CMVI. Physically, this is because in this case the CVE in \eq{lin-b-chiral} (the last term on the right-hand side) can overwhelm the magnetic diffusion (the first term on the right-hand side) and lead to instability. In the presence of rotation $\bO$, Alfv\'en wave is split into two MC waves with frequencies $\o_f$ and $\o_s$. We thus expect that the fast MC wave would become unstable when $|\o_{CA}|>|\o_f|$ and the slow MC wave would become unstable when $|\o_{CA}|>|\o_s|$. Since $|\o_s| < |\o_A|$, we expect that the CMVI is more likely to occur in the rotating case. In the next section, we will show that this is indeed the case.\\
\begin{figure*}[t!]
	\centering
	\includegraphics[width=.490\textwidth]{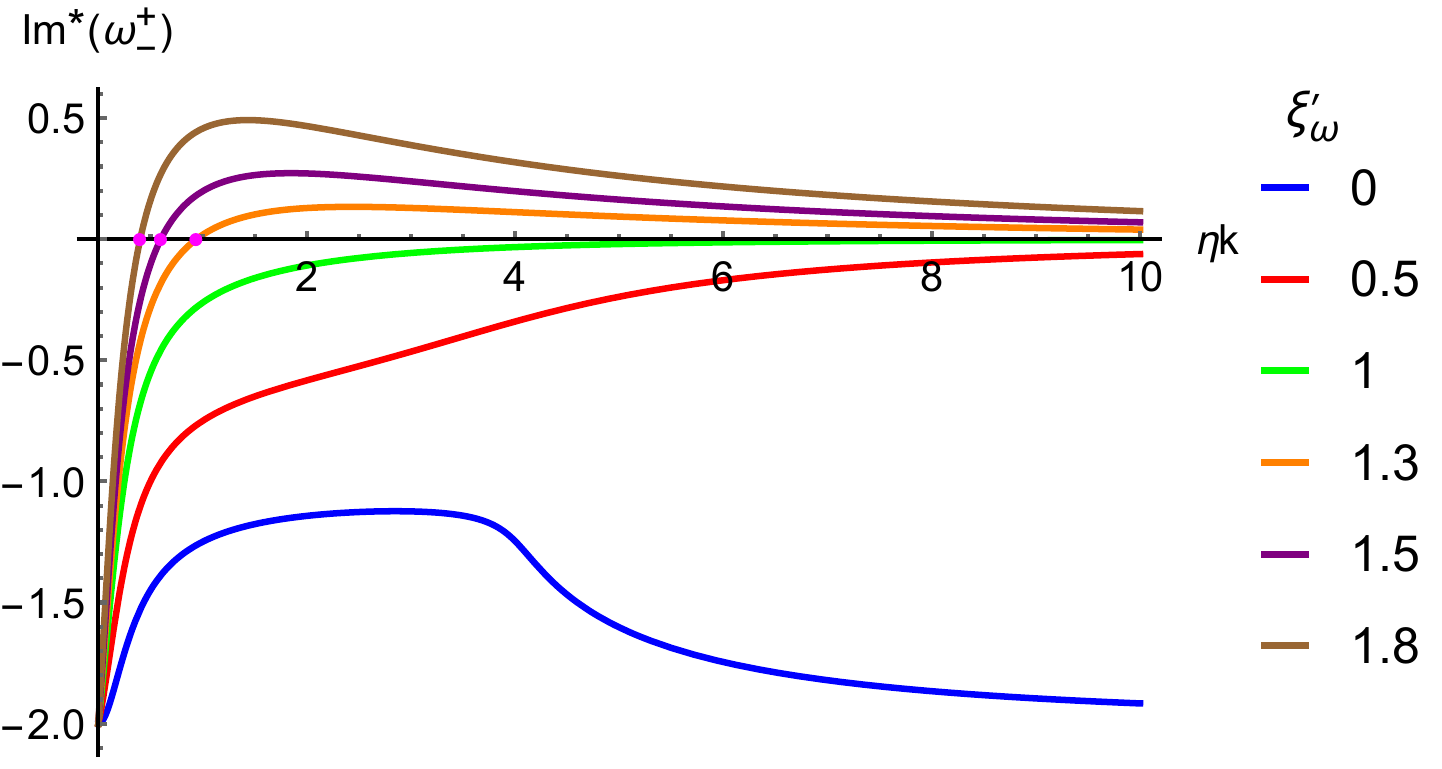}
	\includegraphics[width=.490\textwidth]{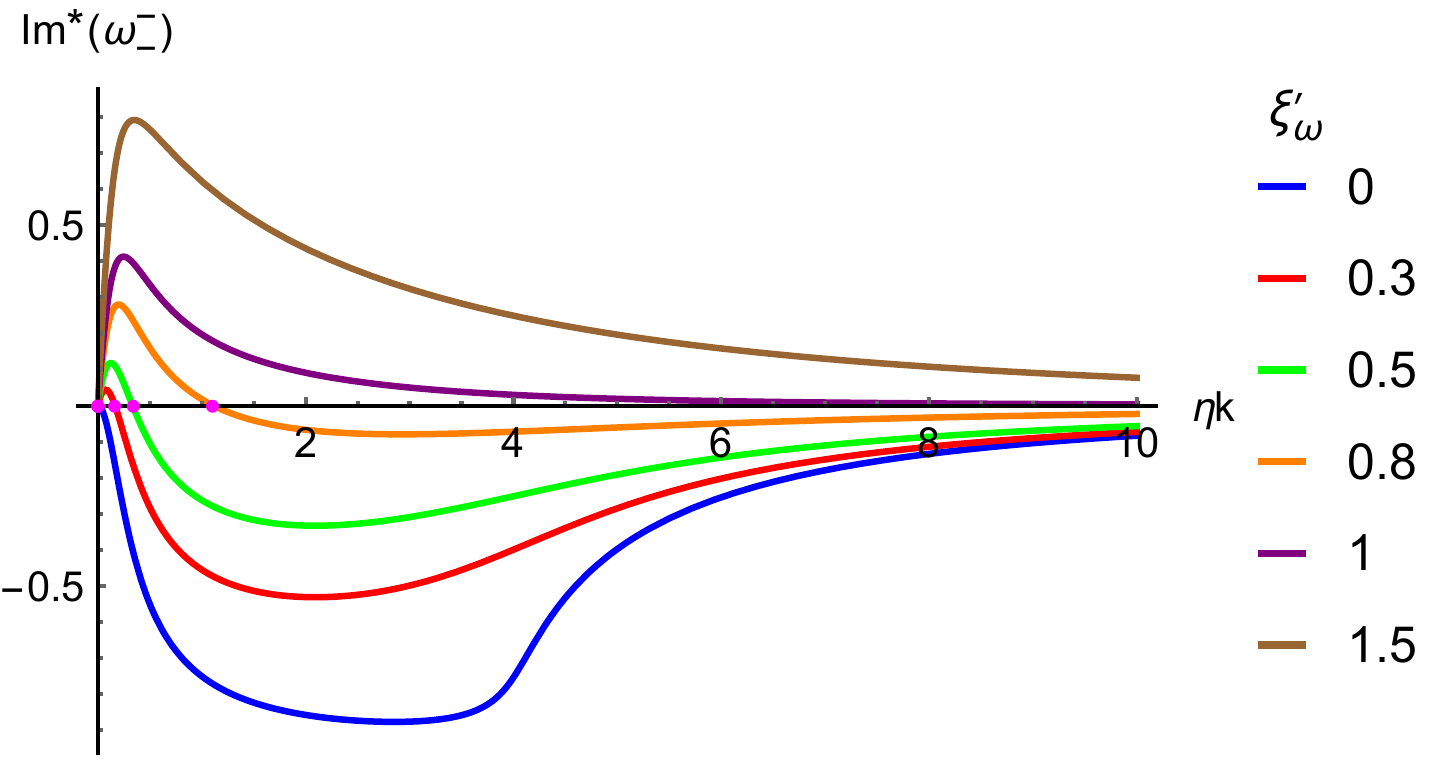}
	
	\caption{
	\textbf{Left:} $\mathrm{Im}^{*}(\o^{+}_{-})\equiv \mathrm{Im}(\o^{+}_{-})/(\g_\w/2)$ as a function of $k$ (in unit of $1/\eta$). Different colors correspond to different CVE coefficients. When $\x'_\o>1$, $\mathrm{Im}^{*}(\o^{+}_{-})$ has a zero point (indicated by the magenta points) located at $2\O\xi_{\o}^{\prime}/B_{0}^{\prime} (\xi_{\o}^{\prime 2}-1)$, which separates the stable and unstable $k$ regimes. 		
		\textbf{Right:} $\mathrm{Im}^{*}(\o^{-}_{-})\equiv \mathrm{Im}(\o^{-}_{-})/(\g_\w/2)$ as a function of $k$. Different colors correspond to different CVE coefficients. When $\x'_\o\geq1$, $\mathrm{Im}^{*}(\o^{-}_{-})$ is always positive. When $\x'_\o<1$, $\mathrm{Im}^{*}(\o^{-}_{-})$ has a zero point (indicated by the magenta points) located at $2\O\xi_{\o}^{\prime}/B_{0}^{\prime} (1-\xi_{\o}^{\prime 2})$, which separates the stable and unstable $k$ regimes. The parameters are $\eta\O=1, B_{0}^{\prime}=4$, and $\h=\pi/3$. } 
	\label{imaginary}
\end{figure*}

\section{CME induced instability}\label{sec:cme}
Considering the CME first ($\x_B\neq 0, \x_\o=0$), \eq{chiral-dispersion-total} simplifies to
\begin{align}
	\label{chiral-dispersion-cmetotal}
	& \o^2+[i\eta k( k +\l \xi_{B})-\l \o_{I} ]\o
   -i\eta k(\xi_{B} +\l k )\o_{I} -\o_A^2  =0.
\end{align}
When $\bO=\bm 0$, it is well known that CME can induce an instability termed chiral plasma instability (CPI) or chiral dynamo instability, which has been extensively studied in the literature~\cite{PhysRevLett.79.1193, PhysRevLett.111.052002,PhysRevD.100.065023}. This instability occurs for the right-hand waves ($\l=-1$) at scales $k<\x_B$. At finite rotation, the solution to \eq{chiral-dispersion-cmetotal} is
\begin{align}
	\o  & = -\frac{1}{2}[i\eta k( k +\l \xi_{B})-\l 
	\o_{I} ]    \nonumber    \\
	&\quad   \pm \frac{1}{2}\sqrt{ 
		[i\eta k( k +\l \xi_{B}) +  \l \o_{I} ]^2
		+ 4 \o_{A}^2  }.  \label{solution-O-B-CME}
\end{align}
When $\eta=0$, Eq.~(\ref{solution-O-B-CME}) reduces to the ideal MHD case. However, even when $\eta\neq0$, for right-hand mode ($\l=-1$), at $k=\xi_{B}$, Eq.~(\ref{solution-O-B-CME}) still yields an ideal-fluid form. This indicates that at this scale, the CME effectively balances the magnetic diffusion. Furthermore, it can be shown (see Appendix \ref{appenanlay}) that for $k>\x_B$, Eq.~(\ref{solution-O-B-CME}) represents only damping waves. For $k<\x_B$, $\l=-1$, the imaginary part of $\o$ is positive and an instability emerges. Therefore, the presence of a finite rotation does not alter the unstable condition $\l=-1, k< \xi_{B}$ for CPI. 

\section{CVE induced instability}\label{sec:cve}
Let us now turn to the dispersion relation when CVE is present ($\x_B= 0, \x_\o\neq0$). From Eq. (\ref{chiral-dispersion-total}), we have 
\begin{equation}
	\o^2+(i \g_{\eta} -\l \o_{I})\o  -\o_{A}^2 
	+i \g_{\eta} (\o_{CA}-\l \o_{I}) = 0 ,
	\label{disper-cve}
\end{equation}
where $\l=+1(-1)$ corresponds to left-hand (right-hand) modes. Denoting $ m= -\g_{\eta}^2 + \o_{I}^2 + 4 \o_{A}^2$ and $ n= 2\g_{\eta}(\l\o_{I} -2\o_{CA})$, the solution to \eq{disper-cve} can be expressed by
\begin{align}
\o=	\o^{\l}_{\chi} & \equiv \frac{1}{2}\big[-i \g_{\eta} + \l \o_{I} + \chi  \sqrt{ m + i\, n} \big] ,     \label{solutom-CVE}
\end{align}
where $\c=\pm 1$. We can directly obtain the real and imaginary parts of $\o^\l_\c$ using the formula
\begin{equation}
	\sqrt{ m +i\, n}= \frac{\sqrt{m+\sqrt{m^2+n^2}}}{\sqrt{2}} + i \, \frac{n}{\sqrt{2}}
	\frac{1}{\sqrt{m+\sqrt{m^2+n^2}}}.  \label{sqrt-m+in}
\end{equation}
The results are 
\begin{align}
	\mathrm{Re}(\o^{\l}_{\chi}) & =  \frac{1}{2}\left[ \l \o_{I} +\chi  \frac{\sqrt{m+\sqrt{m^2+n^2}}}{\sqrt{2}}\right ] , \\
	\mathrm{Im}(\o^{\l}_{\chi}) & = \frac{1}{2}\left[-\g_{\eta} + \chi \frac{n}{\sqrt{2}}
	\frac{1}{\sqrt{m+\sqrt{m^2+n^2}}}\right ]   \nonumber   \\
	& = \frac{ \g_{\eta} }{2}\left[ -1+\chi  
	\frac{\sqrt{2}(\l\o_{I}-2\o_{CA})}{\sqrt{m+\sqrt{m^2+n^2}}} \right] .   \label{Im-CVE}
\end{align}
We focus on $\mathrm{Im}(\o^{\l}_{\chi})$. When $\mathrm{Im}(\o^{\l}_{\chi})>0$, the corresponding mode is unstable. Denoting the angle between $\bB_0$ (and $\bO$) and $\bk$ by $\h$, we have $\o_A=B'_0 k \cos\h$, $\o_I=2\O\cos\h$, and $\o_{CA}=\x'_\o B'_0 k\cos\h$. We restrict our discussions to the case with $\cos\h >0$ so that $\o_A, \o_I, \o_{CA}>0$; the case with $\cos\h<0$ can be similar studied. Through direct calculation, we find that when $\c=+1$, $\mathrm{Im}(\o^{\l}_{+})<0$ ($\l=\pm 1$) for all $k>0$ and all parameter regions; hence, they correspond to damped wave modes. For $\chi =-1$, we find that there exist kinematic and parameter regions such that $\mathrm{Im}(\o^{\l}_{-})>0$, leading to instability:
\begin{align}
	& \mathrm{Im}(\o^{+}_{-})>0 \Leftarrow   
	\o_{CA}^{2}-\o_{I}  \o_{CA} > \o_{A}^2 , \label{o+--1} \\
	&\mathrm{Im}{(\o^{-}_{-})}>0 \Leftarrow  
	\o_{CA}^{2}  + \o_{I} \o_{CA} >\o_{A}^2 .  \label{o---1}
\end{align}
It is instructive to equivalently rewrite the instability conditions as
\begin{align}
	& \mathrm{Im}(\o^{+}_{-})>0 \quad \Leftarrow  \quad 
	\o_{CA} > \o_{f} , \label{o+--122} \\
	&\mathrm{Im}{(\o^{-}_{-})}>0 \quad \Leftarrow \quad 
	\o_{CA} >\o_{s} .  \label{o---1222}
\end{align}
These conditions are precisely those anticipated from the discussion in Sec.~\ref{sec:cmhd}. It is interesting to notice that these conditions do not depend on $\cos\h$ as long as $\cos\h>0$.

In Fig.~\ref{imaginary}, we plot the normalized imaginary parts of the frequencies, $\mathrm{Im}^*(\o^{+}_{-}) \equiv \mathrm{Im}(\o^{+}_{-})/(\g_\w/2)$ and $ \mathrm{Im}^*(\o^{-}_{-})\equiv\mathrm{Im}(\o^{-}_{-})/(\g_\w/2)$, as functions of $k$ (in unit of $1/\w$), for the case of $\cos\h >0$. The $\o^{+}_{-}$ mode exhibits instability for $1<\xi_{\o}^{\prime}$ and for $k$ exceeding certain threshold $k>2\O\xi_{\o}^{\prime}/B_{0}^{\prime} (\xi_{\o}^{\prime 2}-1)$, which can be verified analytically from condition (\ref{o+--122})), and the threshold value decreases when $\xi_{\o}^{\prime}$ increases. For $\o^{-}_{-}$, two distinct scenarios arise. When $1 \leq \xi_{\o}^{\prime}$, all modes with $0<k$ are unstable. Conversely, when $0<\xi_{\o}^{\prime}<1$, 
the instability is confined only to small $k$ values, with an upper bound given by $2\O\xi_{\o}^{\prime} / B_{0}^{\prime} (1-\xi_{\o}^{\prime 2})$ ( which can be deduced directly from the condition (\ref{o---1222})). 

We can perform a similar analysis for the case of $\cos\h <0$. In Appendix \ref{appenanlay}, we provide an additional detailed analysis utilizing a geometric approach, which yields the same results for both $\cos\h>0$ and $\cos\h<0$ cases. For clarity, we summarize the results of the instability analysis in Table \ref{tab-cve+case12} to facilitate the identification of regions where instabilities occur. The results show that the rotation can dramatically catalyze the onset of CMVI. This is most evident for mode with $\c=-1, \l=-1$ ($\c=+1, \l=-1$) for $\cos\h>0$ ($\cos\h <0$) case. In this situation, the CMVI emerges for any finite $\x_\o^\prime$, which makes it particularly  relevant for realistic rotating chiral plasmas, as $\x'_\o$ is typically expected to be small.  
\begin{table}[h!]
		\centering
		\begin{tabular}{ | c | c | c |} 	
			\hline  
			 & $\x'_\o$ & $ k\; {\rm and}\,\cos\h $  \\[0.5ex]
			\hline & &\\[-2.5ex] 
	\,\,	$\mathrm{Im}(\o_+^+)>0$ \quad &  $\displaystyle\xi_{\o}^{\prime} > 1$  &\,  $\displaystyle \cos\h <0,\, k>\frac{2\xi_{\o}\O}{B_{0}(\xi_{\o}^{\prime2}-1)} $ \\[2ex]
				\hline
		\multirow{2}{*}{$\mathrm{Im}(\o_+^-)>0$} & 	$\displaystyle\xi_{\o}^{\prime} \geq 1$ & $\cos\h<0,\, k>0$ \\[0.3ex]
\cline{2-3}& &\\[-2.5ex]
& \, $1>\xi_{\o}^{\prime} > 0$ \,\quad& \, $\displaystyle\cos\h<0,\, 0<k<\frac{2\xi_{\o}\O}{B_{0}(1-\xi_{\o}^{\prime2})}$\, \\[2ex]			\hline & &\\[-2.5ex]
			$\mathrm{Im}(\o_-^+)>0$ &  $\xi_{\o}^{\prime} > 1$  &  $\displaystyle\cos\h >0,\, k>\frac{2\xi_{\o}\O}{B_{0}(\xi_{\o}^{\prime2}-1)} $ \\[2ex]
			\hline
		\multirow{2}{*}{$\mathrm{Im}(\o_-^-)>0$} & 	$\xi_{\o}^{\prime} \geq 1$ & $\cos\h>0,\, k>0$ \\[0.3ex]
		\cline{2-3}& &\\[-2.5ex]
		& $1>\xi_{\o}^{\prime} > 0$ & $\displaystyle\cos\h>0,\, 0<k<\frac{2\xi_{\o}\O}{B_{0}(1-\xi_{\o}^{\prime2})}$ \\ [2ex]
			\hline
		\end{tabular}
				\caption{ Conditions for CMVI under rotation for different $\x'_\o$ and different kinematic regions. Note that these conditions depend only on the sign of $\cos\h$.}	\label{tab-cve+case12}
	\end{table}
\begin{figure*}[!thb]
\begin{center}
\hspace*{-10mm}
\includegraphics[clip, width=0.45\textwidth]{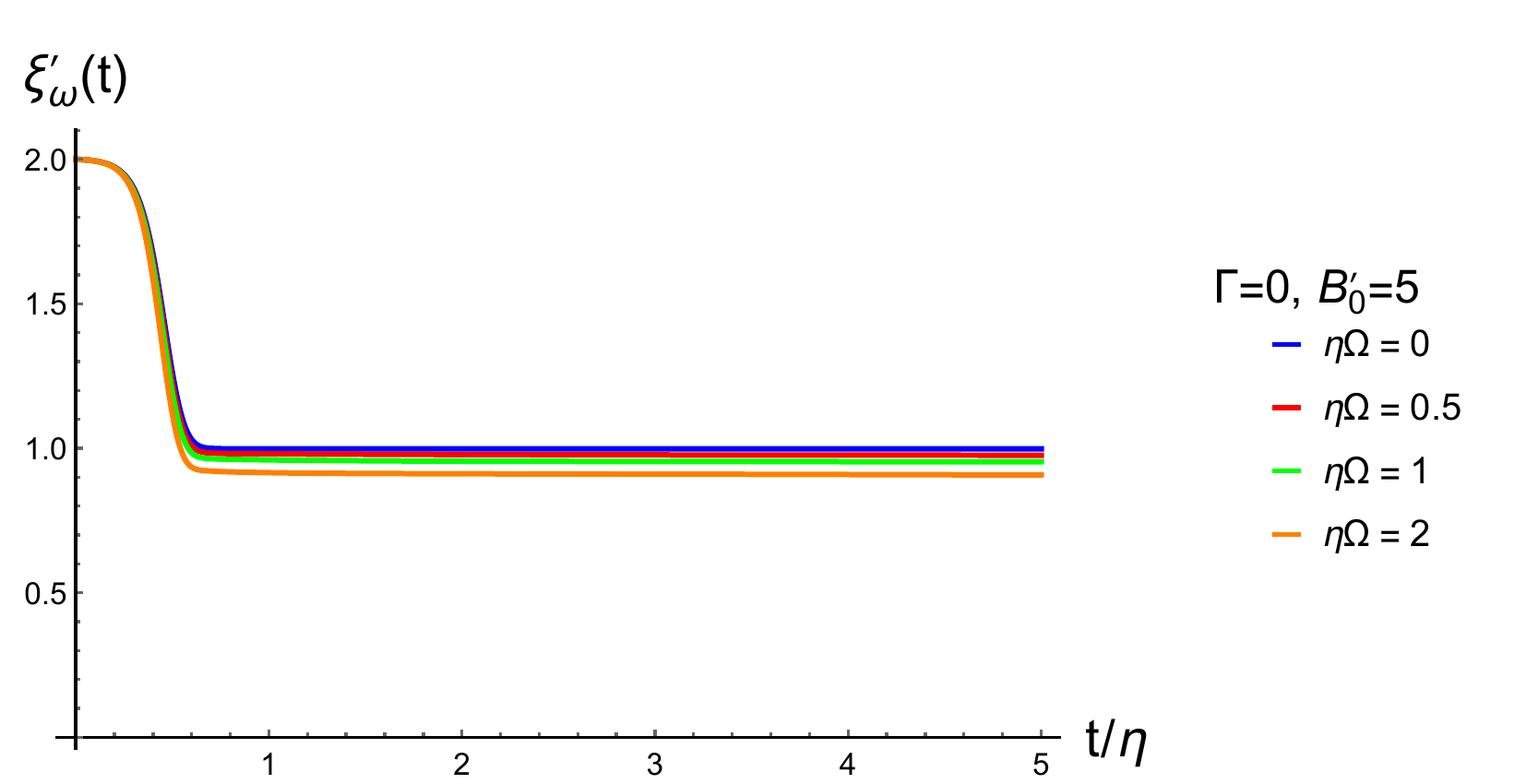}
\hspace*{5mm}
\includegraphics[clip, width=0.45\textwidth]{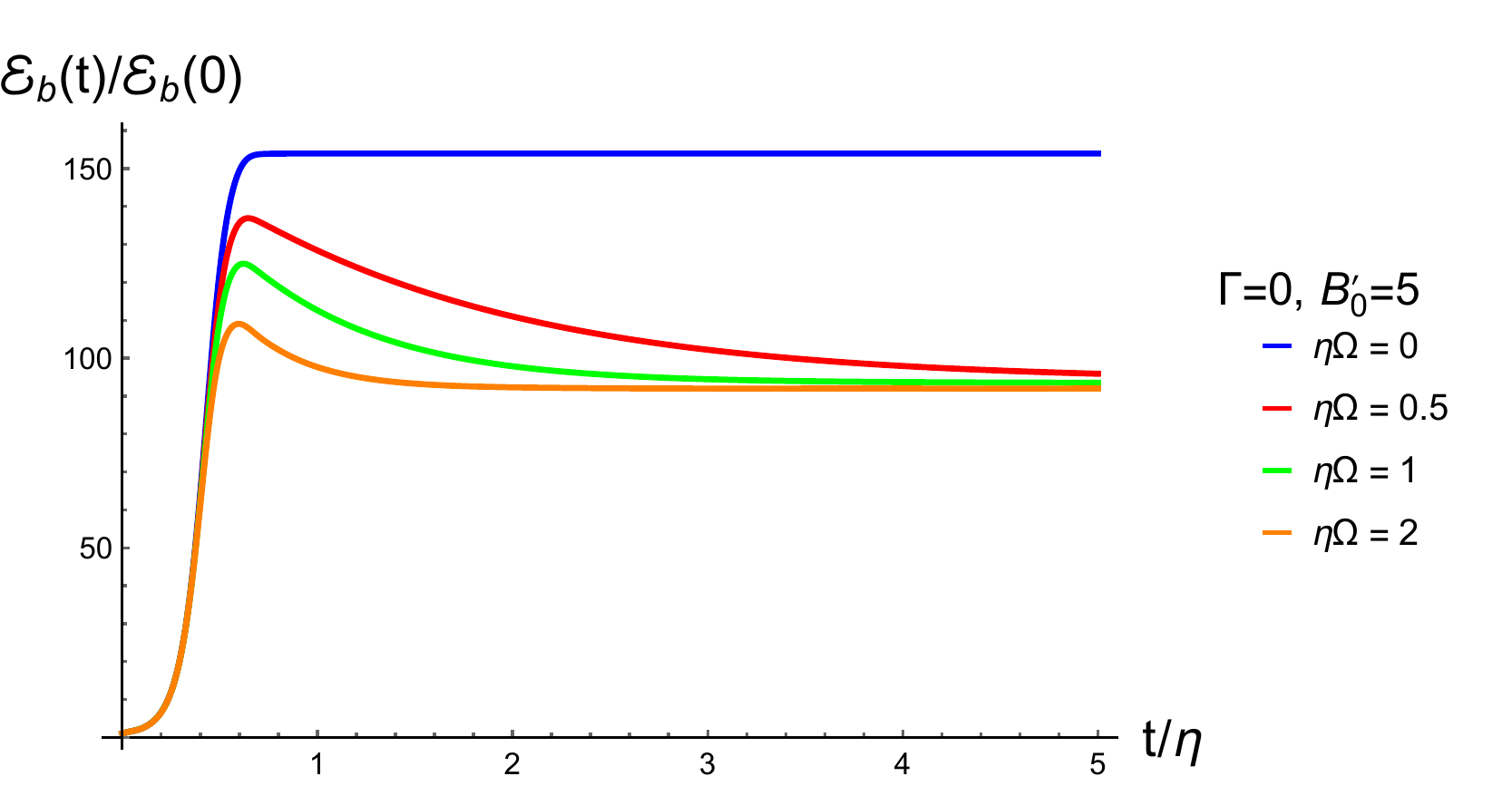}
\vspace*{5mm}\hfill \\
\mbox{(I) \textbf{Left:} The evolution of CVE coefficient $\xi_{\o}^{\prime}(t)$.  \textbf{Right:} The evolution of magnetic energy $\mathcal{E}_{b}(t)/\mathcal{E}_{b}(0)$ } \hfill \\
\vspace*{5mm}
\hspace*{-10mm}
\includegraphics[clip, width=0.45\textwidth]{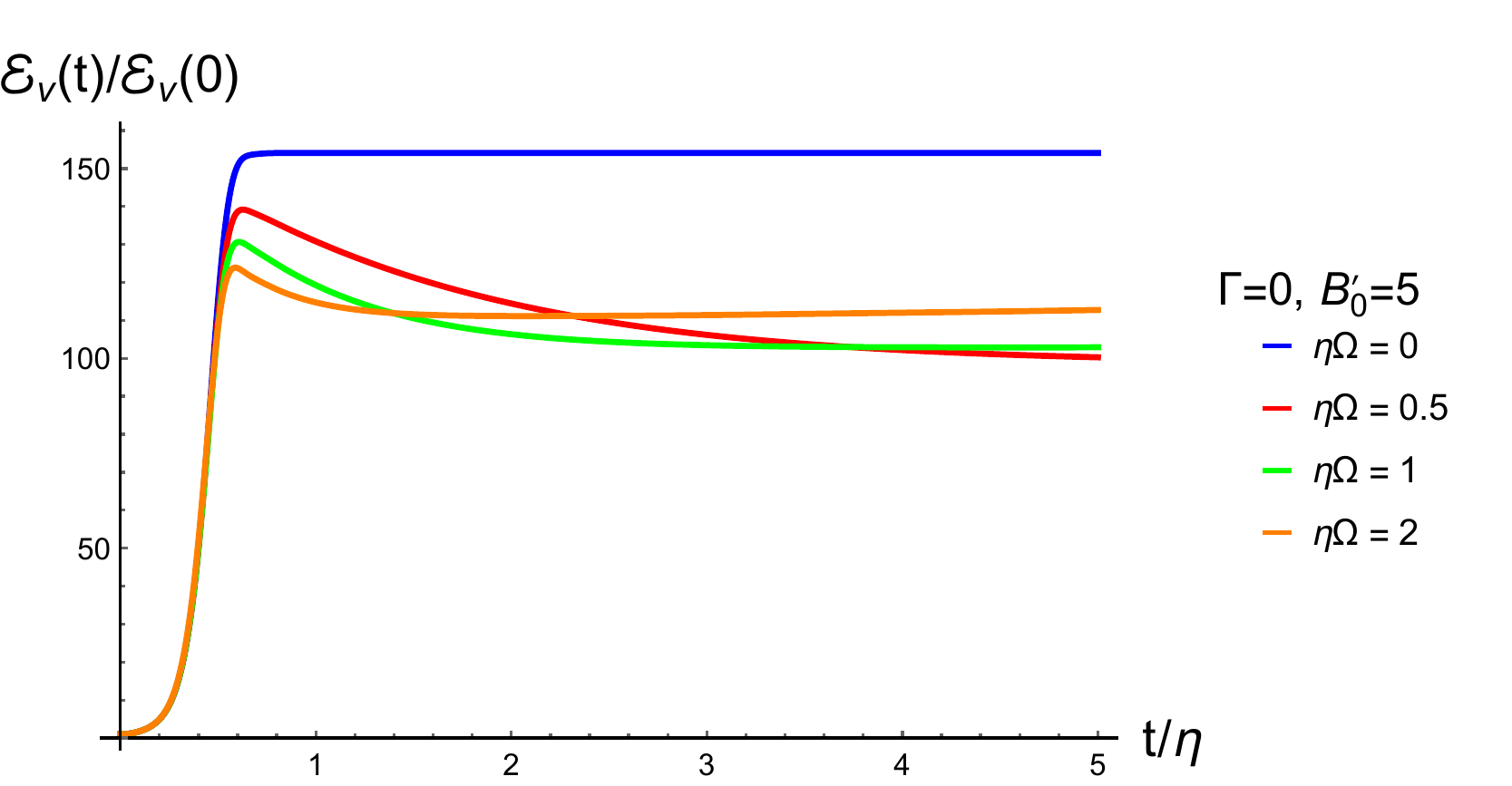}
\hspace*{5mm}
\includegraphics[clip, width=0.45\textwidth]{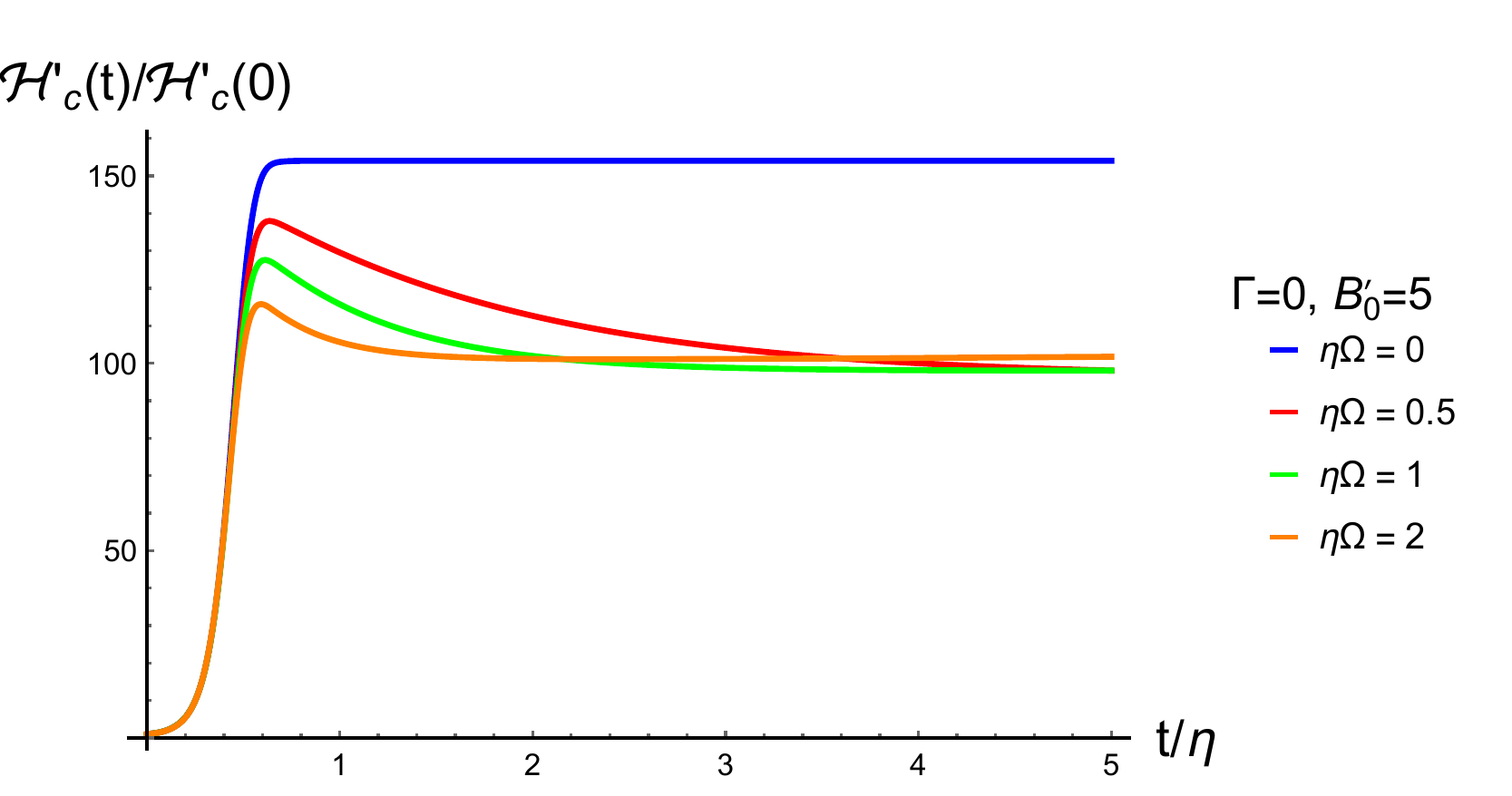}
\vspace*{5mm}\hfill \\
\mbox{(II) \textbf{Left:} The evolution of kinetic energy $\mathcal{E}_{v}(t)/\mathcal{E}_{v}(0)$. \textbf{Right:} The evolution of cross helicity $\mathcal{H}'_{c}(t)/\mathcal{H}'_{c}(0)$. } \hfill \\
\vspace*{5mm}
\hspace*{-10mm}
\includegraphics[clip, width=0.45\textwidth]{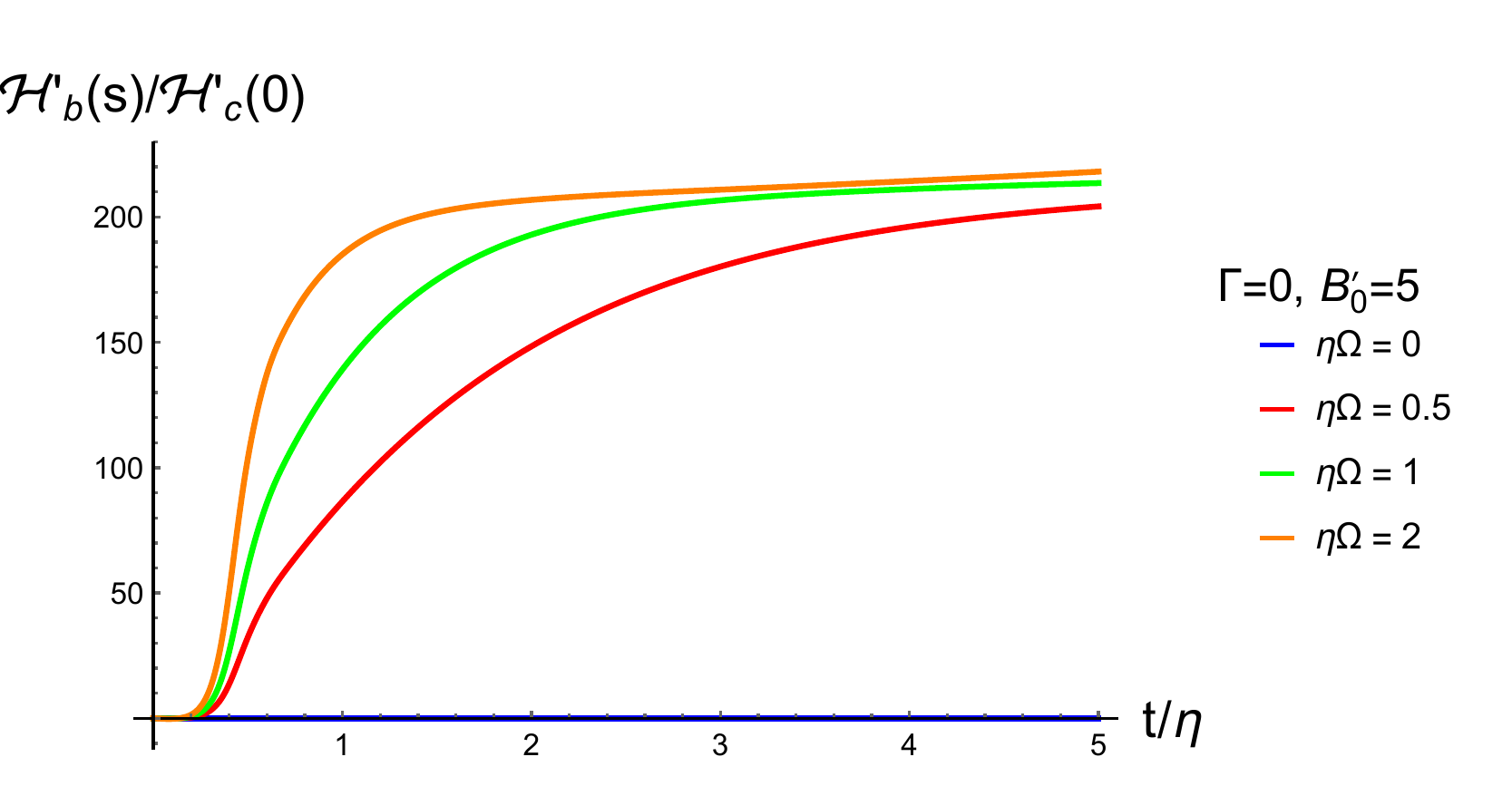}
\hspace*{5mm}
\includegraphics[clip, width=0.45\textwidth]{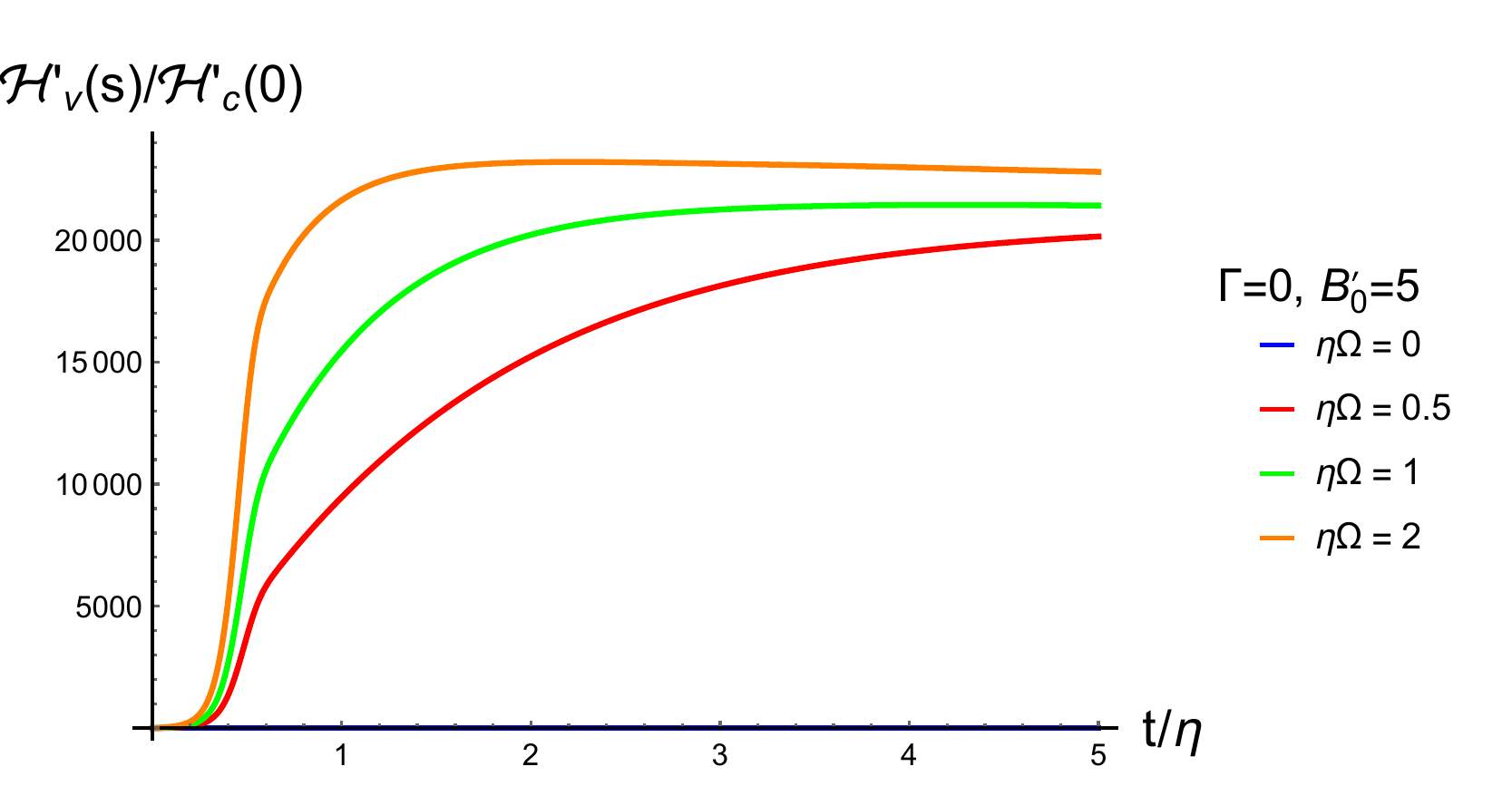}
\vspace*{5mm} \hfill \\
\mbox{(III) \textbf{Left:} The evolution of magnetic helicity $\mathcal{H}'_{b}(t)/\mathcal{H}'_{c}(0)$. \textbf{Right:} The evolution of kinetic helicity $\mathcal{H}'_{v}(t)/\mathcal{H}'_{c}(0)$. }
\caption{\label{fig-big-ome} The numerical results of the normalized CVE coefficient, magnetic energy, kinetic energy, cross helicity, magnetic helicity, and kinetic helicity as a function of time $t$. Different colors correspond to different rotations. The initial value for CVE coefficient is $\xi'_\o(0)=2$.  }
\end{center}
\end{figure*}
\begin{figure*}[!thb]
\begin{center}
\includegraphics[clip, width=0.45\textwidth]{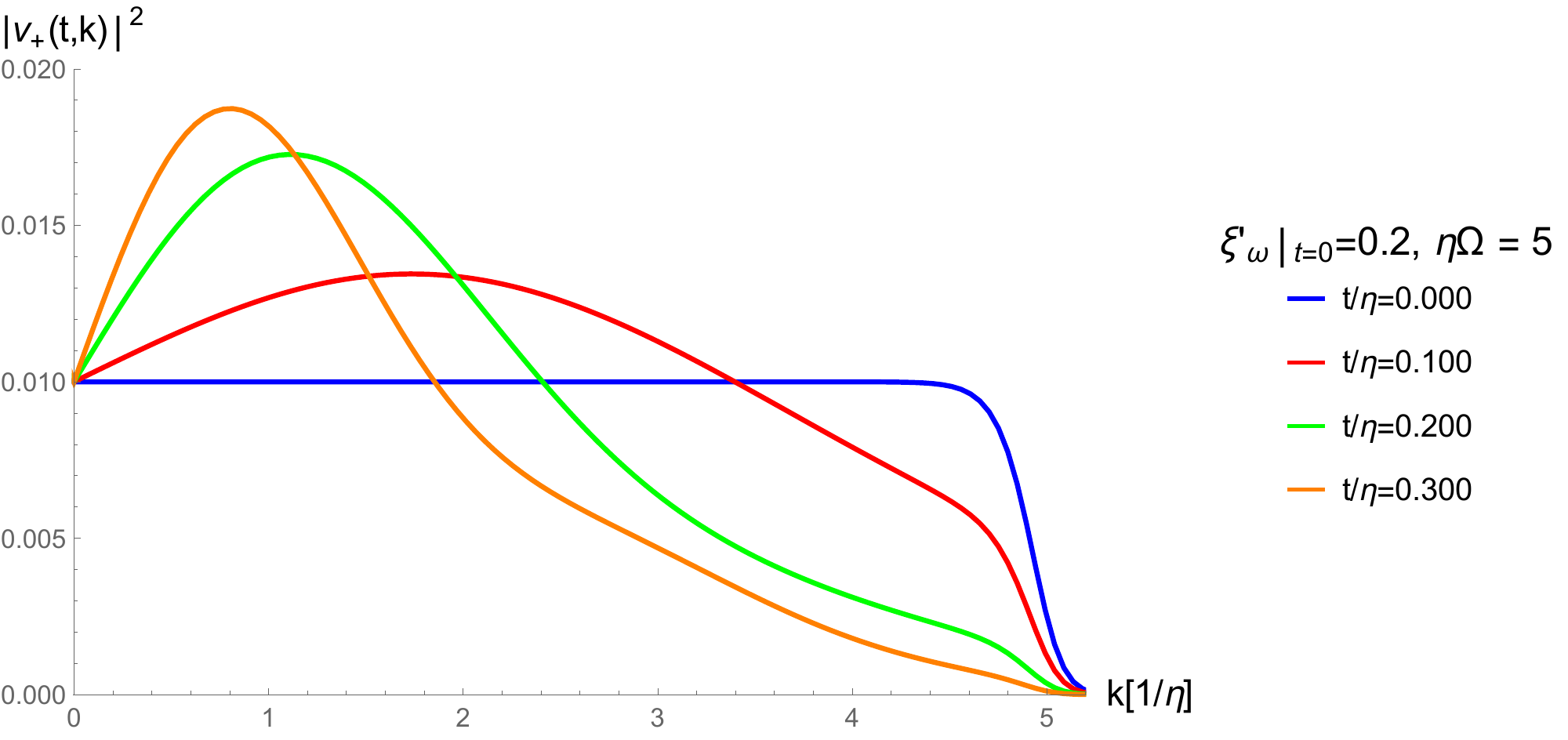}
\hspace*{5mm}
\includegraphics[clip, width=0.45\textwidth]{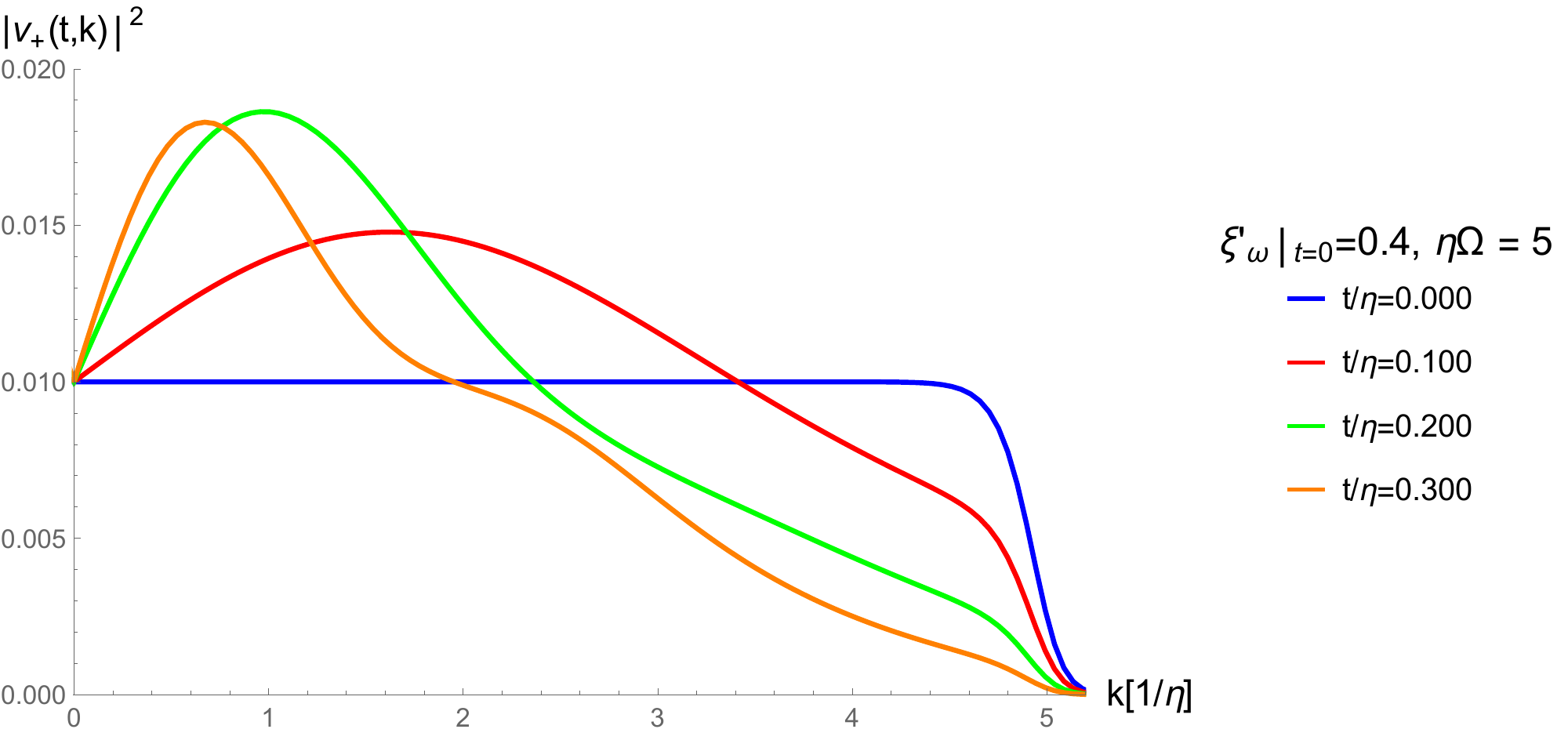}
\vspace*{5mm}\hfill \\
\includegraphics[clip, width=0.45\textwidth]{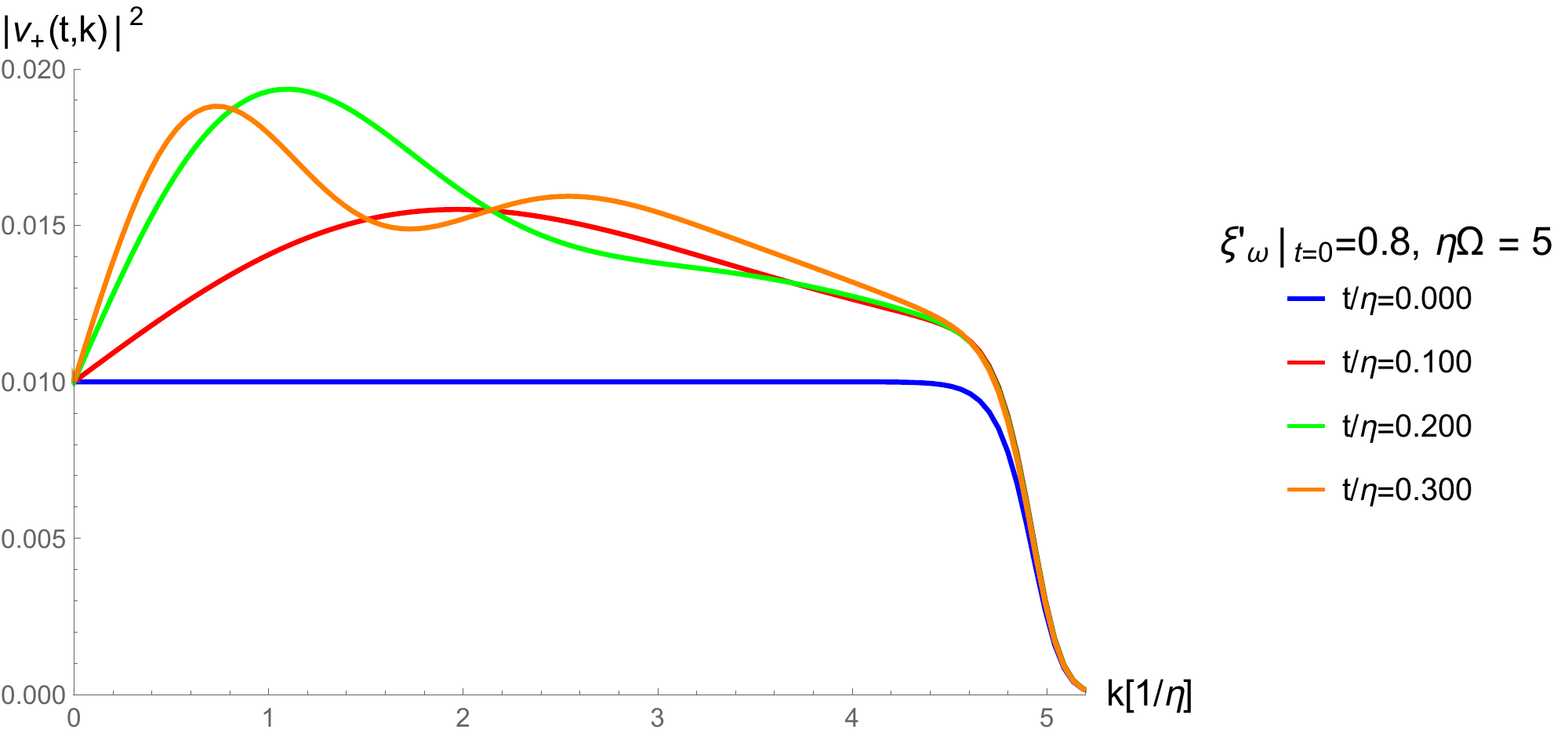}
\hspace*{5mm}
\includegraphics[clip, width=0.45\textwidth]{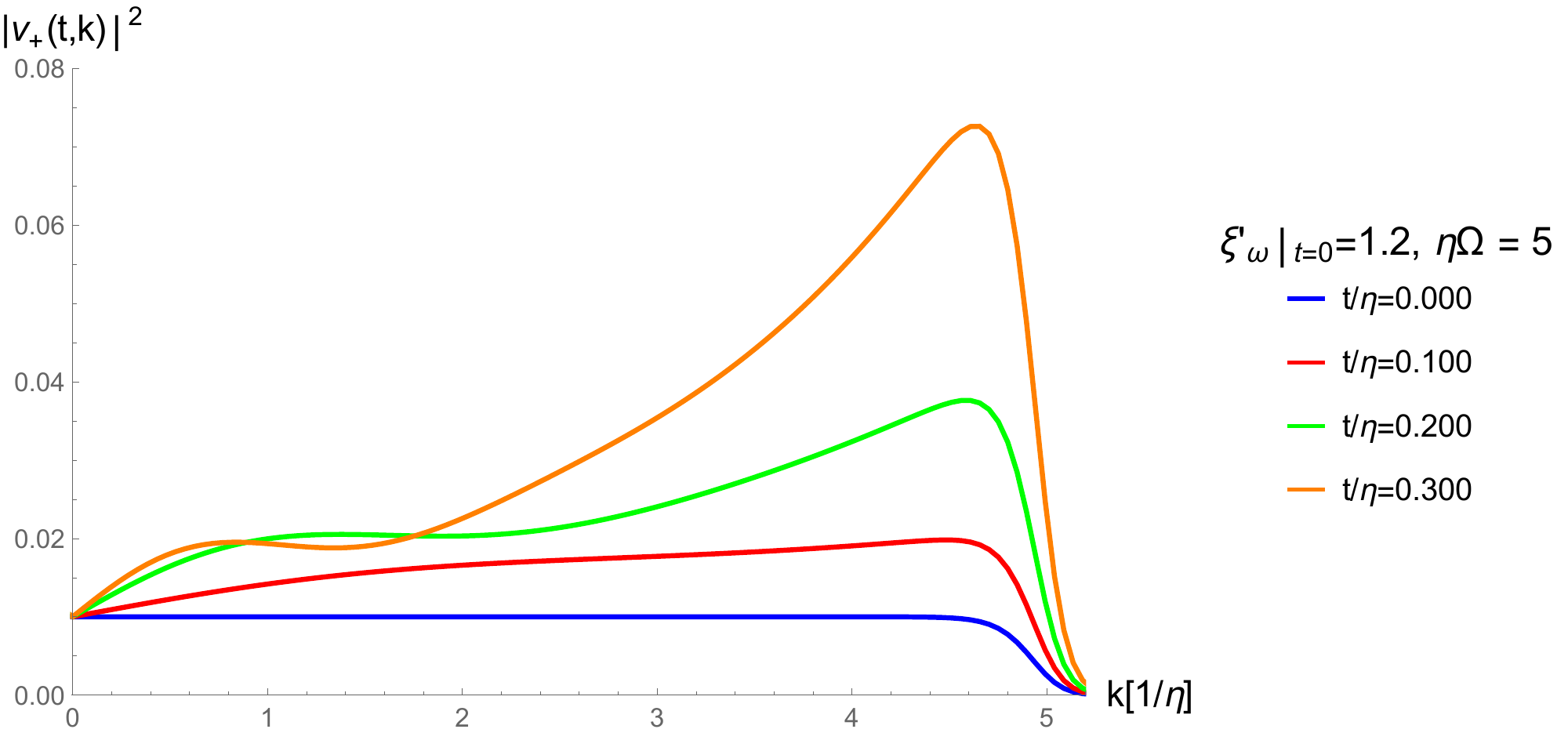}
\caption{\label{fig-vpo-t} Square of the right-handed velocity component, $|v_+(t,\bk)|^2$, in units of $\eta^6$ for different initial values of $\x'_\o=0.2, 0.4, 0.8,$ and $1.2$ (from top left to bottom right). The other parameters are $\G=0$, $\O=5/\eta$, and $B'_0=5$. Different colors correspond to different times. }
\end{center}
\end{figure*}
\begin{figure*}[!thb]
\begin{center}
\hspace*{-10mm}
\includegraphics[clip, width=0.45\textwidth]{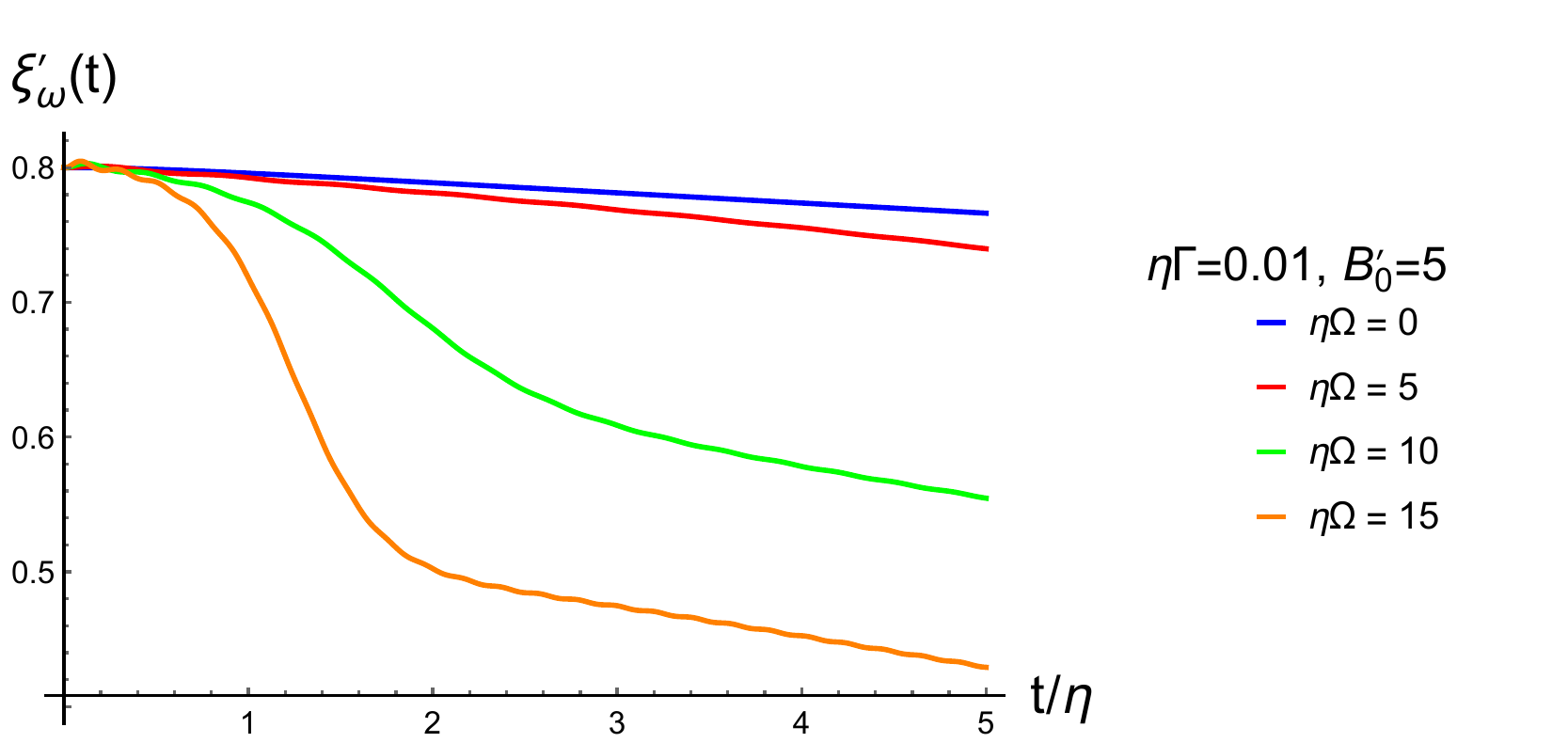}
\hspace*{5mm}
\includegraphics[clip, width=0.45\textwidth]{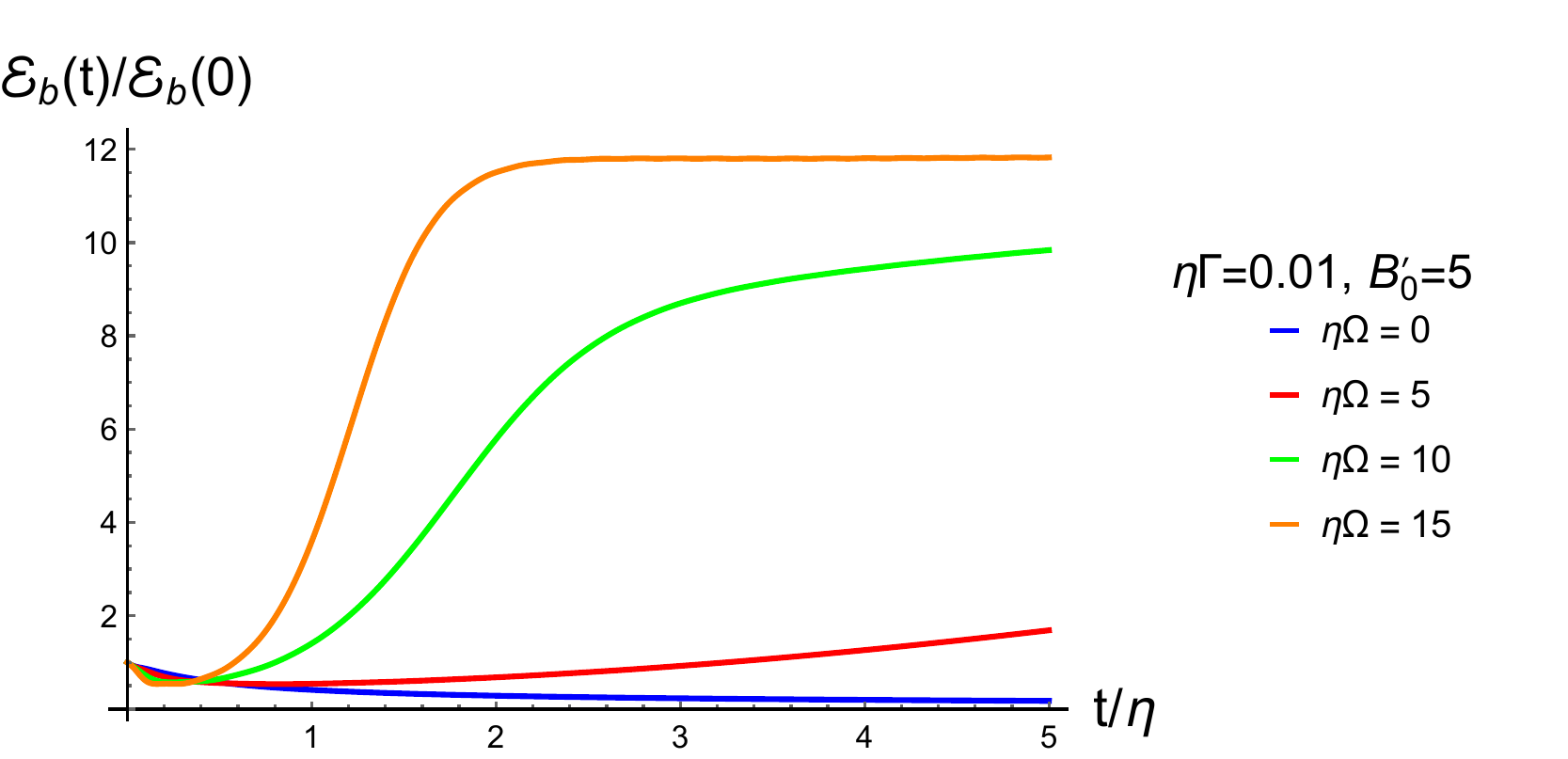}
\vspace*{5mm}\hfill \\
\mbox{(I) \textbf{Left:} The evolution of CVE coefficient $\xi_{\o}^{\prime}(t)$.  \textbf{Right:} The evolution of magnetic energy $\mathcal{E}_{b}(t)/\mathcal{E}_{b}(0)$ } \hfill \\
\vspace*{5mm}
\hspace*{-10mm}
\includegraphics[clip, width=0.45\textwidth]{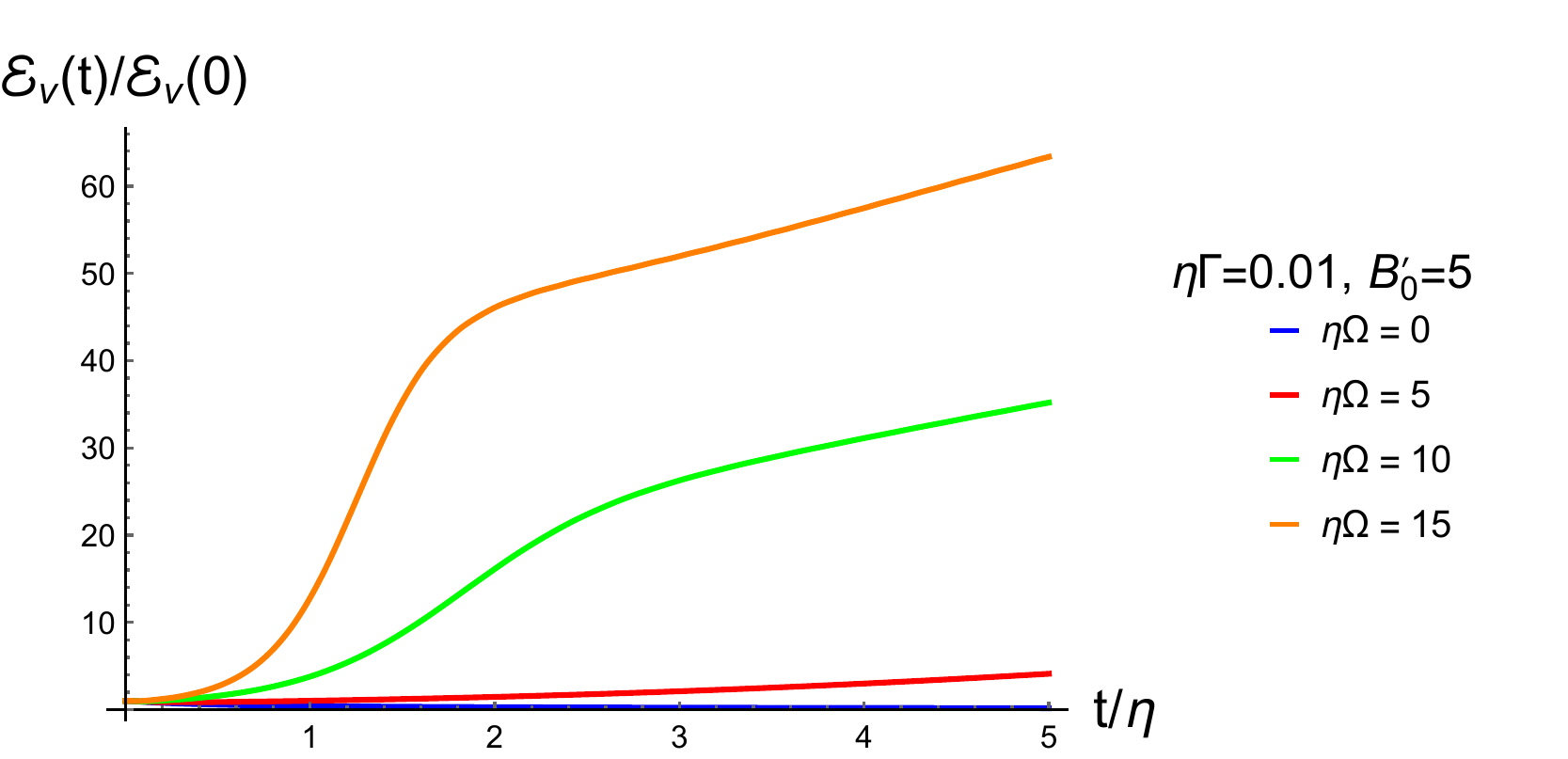}
\hspace*{5mm}
\includegraphics[clip, width=0.45\textwidth]{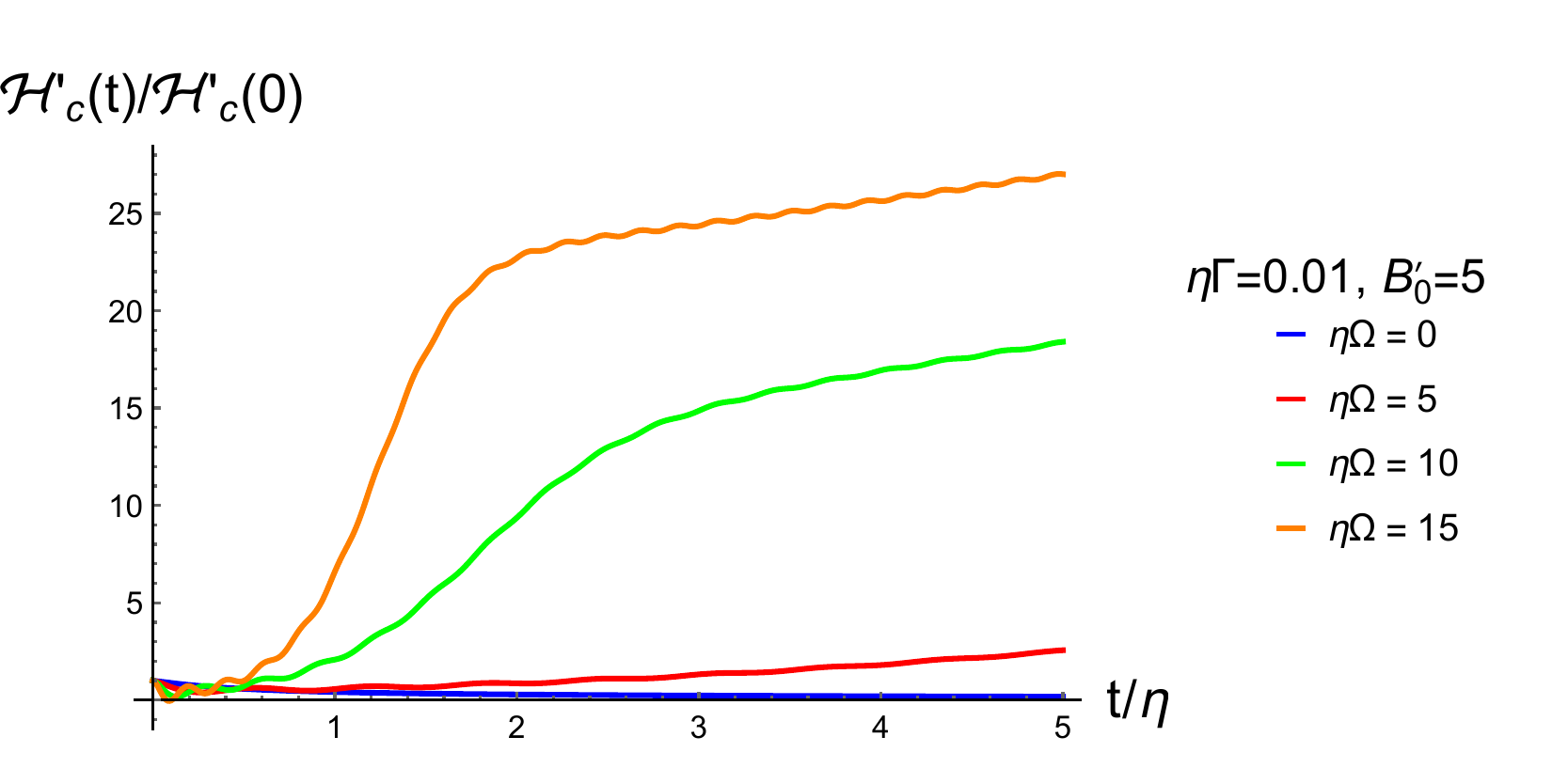}
\vspace*{5mm}\hfill \\
\mbox{(II) \textbf{Left:} The evolution of kinetic energy $\mathcal{E}_{v}(t)/\mathcal{E}_{v}(0)$. \textbf{Right:} The evolution of cross helicity $\mathcal{H}'_{c}(t)/\mathcal{H}'_{c}(0)$. } \hfill \\
\vspace*{5mm}
\hspace*{-10mm}
\includegraphics[clip, width=0.45\textwidth]{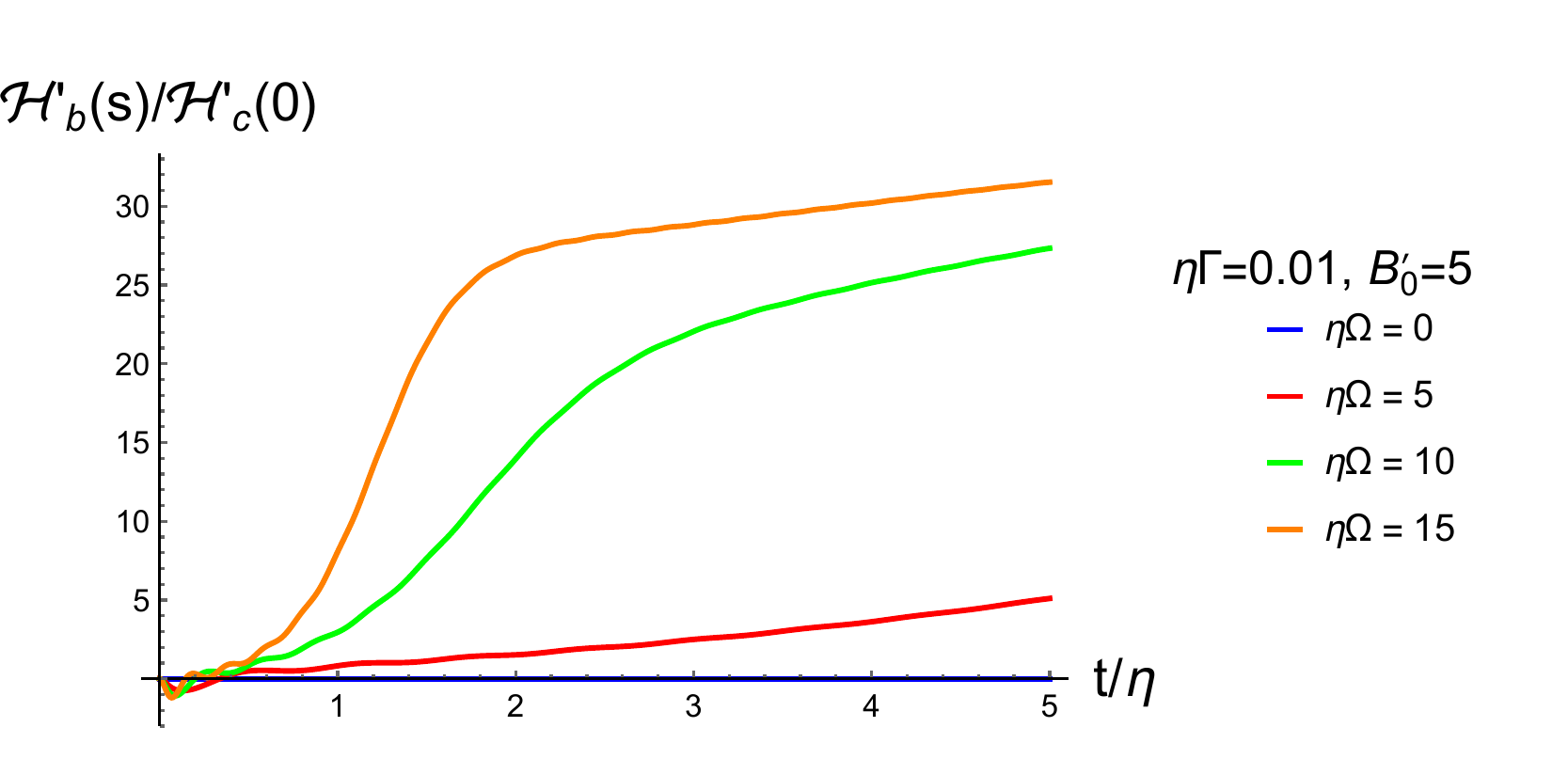}
\hspace*{5mm}
\includegraphics[clip, width=0.45\textwidth]{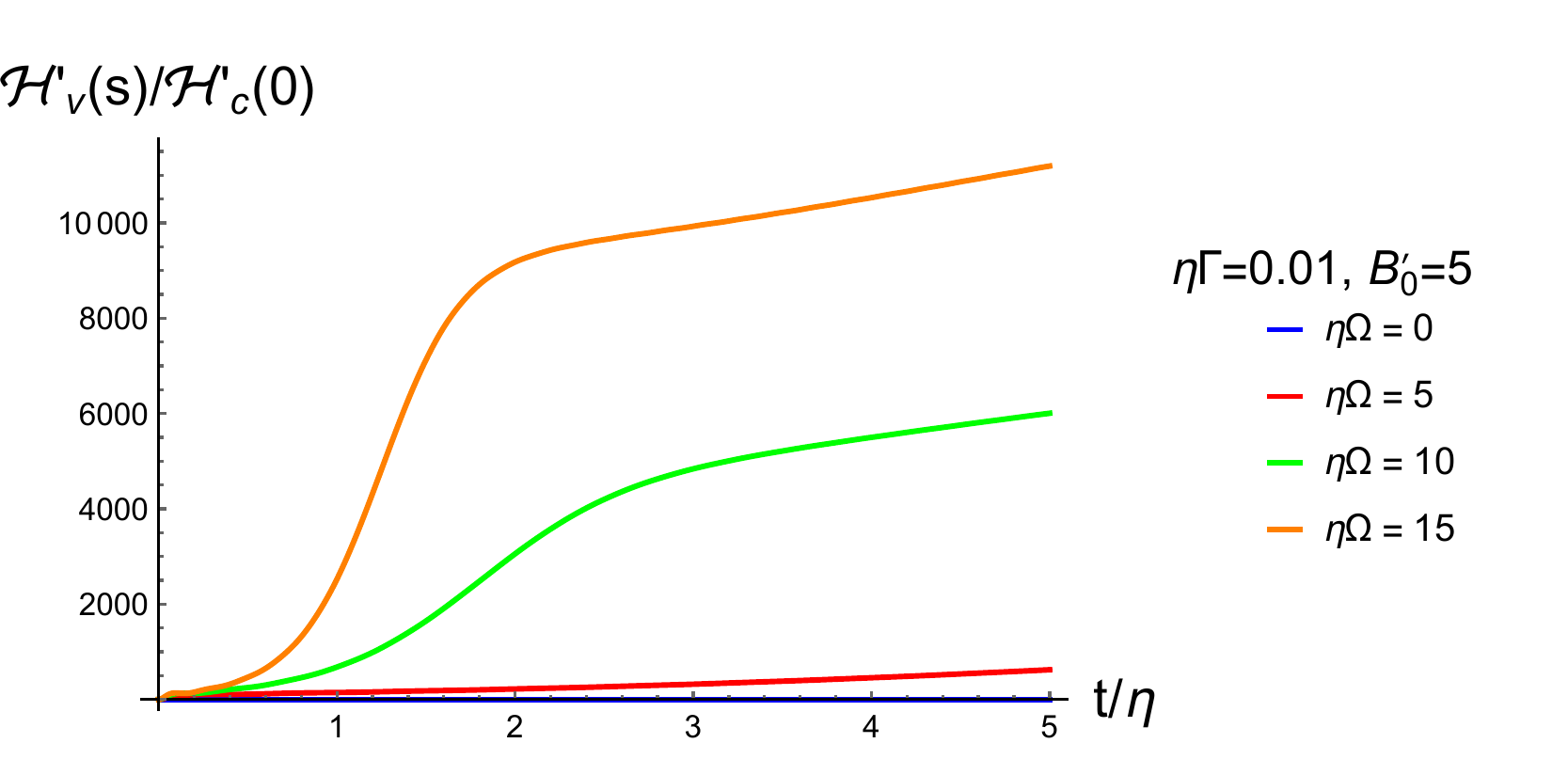}
\vspace*{5mm} \hfill \\
\mbox{(III) \textbf{Left:} The evolution of magnetic helicity $\mathcal{H}'_{b}(t)/\mathcal{H}'_{c}(0)$. \textbf{Right:} The evolution of kinetic helicity $\mathcal{H}'_{v}(t)/\mathcal{H}'_{c}(0)$. }
\caption{\label{fig-diff-omega} The numerical results of the normalized CVE coefficient, magnetic energy, kinetic energy, cross helicity, magnetic helicity, and kinetic helicity as a function of time $t$. Different colors correspond to different rotations. The initial value for CVE coefficient is $\xi'_\o(0)=0.8$ and the chirality flipping rate is $\G=0.01/\eta$.  }
\end{center}
\end{figure*}

\section{Evolution of the instability}\label{sec:evolu}
In a realistic plasma, the instability cannot persist forever, there must be other physical processes to terminate the instability eventually. In order to study the fate of the rotation catalyzed CMVI, similar to the discussion in Ref.~\cite{PhysRevD.109.L121302}, we examine the chiral anomaly equation (see Appendix \ref{derivofeqns} for a derivation)
\begin{align}
	\pt_{t} j_{5}^0 + \bnabla \cdot \bj_{5} = C \bE \cdot \bB,
	\label{chiral-anom-3D}
\end{align}
where $C$ is a constant representing the strength of chiral anomaly,  and (see Appendix \ref{derivofeqns} for a derivation):
\begin{align}
	j^{0}_{5} & \approx  n_{5} + \k_{B} \bB\cdot(\bv+\bu_{\O}) + \k_{\o} ( \bo + 2 \bO)\cdot (\bv + \bu_{\O} ),    \\
	\bj_{5} & \approx n_{5}\bv + \k_{B} \bB + \k_{\o}(\bo+2\bO),
\end{align}
with $n_{5}$ the chiral charge density of constituent particles, $\k_{B}\propto \m$ (with $\m$ the vector chemical potential) the coefficient of the chiral separation effect, and $\k_{\o}\propto T^2$ characterizing the axial CVE. To proceed, we introduce the Fourier modes of velocity and magnetic fields,
\begin{align}
\bv(t,\bx)&=\int_{\bk}\bv(t,\bk)e^{i\bk\cdot \bx}, \label{fourie-v} \\
\bb(t,\bx)&=\int_{\bk}\bb(t,\bk) e^{i\bk\cdot\bx}, \label{fourie-b}
\end{align}
where $\int_{\bk}\equiv\int d^{3}\bk/(2\pi)^3$. The conditions $\bnabla\cdot\bv=\bnabla\cdot\bb=0$ imply that $\bv \ (t,\bk)=v_{+}(t,\bk)\be_{+}(\bk)+v_{-}(t,\bk)\be_{-}(\bk)$ and $\bb \ (t,\bk)=b_{+}(t,\bk)\be_{+}(\bk)+b_{-}(t,\bk)\be_{-}(\bk)$. Then \eq{linear-v} and \eq{lin-b-chiral} lead to 
\begin{align}
\pt_{t}v_{\pm} & = \pm\frac{2i\bO\cdot \bk}{k} v_{\pm} + \frac{i\bB_{0}\cdot\bk}{\r} b_{\pm}, \label{spectrum-v} \\
\pt_{t}b_{\pm} &  = (i\bB_{0}\cdot \bk+\eta \bk^2 \xi_{\o})  v_{\pm} -\eta\bk^2\ b_{\pm} . \label{spectrum-b}
\end{align}
When $\xi_{\o}$ is a constant, \eqs{spectrum-v}{spectrum-b} can be solved analytically, as detailed in Appendix \ref{spectrum-analysis}, which offer a direct pathway to analyze the evolution of magnetic energy, kinetic energy, and various helicities (see definitions below). When $\xi_\o$ is time dependent, \eqs{spectrum-v}{spectrum-b} should be solved in couple with \eq{chiral-anom-3D},  to determine the fate of the rotation-catalyzed CMVI. 

Assuming a homogeneity of the system and writing $n_{5}(t)=\chi_{5}\m_{5}(t)$ with $\chi_{5}\propto T^2$ the chiral susceptibility, we obtain the following evolution equation from \eq{chiral-anom-3D}: 
\begin{align}
\chi_{5} \pt_{t}\m_{5} = - \k_{B} \pt_{t}\mathcal{H}_{c}-\k_{\o}\pt_{t}\mathcal{H}_{v}-\frac{C}{2}\pt_{t}\mathcal{H}_{b}- \G \chi_{5} \m_{5}, \label{dmu5/dt}
\end{align}
where we have added a chirality-flipping term (the last term on the right-hand side) originated from, e.g., the finite mass of the particles~\cite{Grabowska:2014efa}. In Eq.~(\ref{dmu5/dt}), the terms $\mathcal{H}_{c}= \langle (\bv+\bu_{\O})\cdot \bB \rangle$, $\mathcal{H}_{v}=\lan (\bv+\bu_{\O})\cdot (\bo+2\bO) \ran$, and $\mathcal{H}_{b}=\lan \bm{A}\cdot\bB\ran$ represent the averaged cross, kinetic, and magnetic helicities, respectively, with $\lan...\ran\equiv V^{-1}\int d^3\bx(...)$. Therefore, when the chirality relaxation rete $\G=0$, Eq.~(\ref{dmu5/dt}) implies a generalized helicity conservation relation~\cite{Kirilin:2017tdh}. Let $\bm{A}= \bm{A}_0+\bm{a}$ where $\bm{A}_0=\bB_{0}\times\bx/2$ is the vector potential for background magnetic field and $\bm{a}$ is the fluctuating component of the vector potential such that $\bb=\bnabla\times\bm{a}$. 

In our calculation, we have assumed $\bB_{0}\parallel\bO$ which ensures $\bB_{0}\cdot\bu_{\O} =0$. Furthermore, we consider fluctuations $\bv$ and $\bb$ to be transverse to $\bB_0$ and to vary mainly along $\bB_0$ direction. Under these setups, as we show in Appendix \ref{helicity-analysis}, one finds that $\lan\bB_{0}\cdot\bv\ran=\lan\bO\cdot\bv\ran=\lan\bu_\O\cdot\bb\ran=\lan\bm{A}_0\cdot\bb\ran=\lan\bm{a}_0\cdot\bB_0\ran=\lan\bu_\O\cdot\bo\ran=0$. Therefore, we have 
\begin{align}
\mathcal{H}_{c}(t)=\lan\bv\cdot \bb\ran, \
\mathcal{H}_{v}(t)=\lan\bv\cdot\bo\ran,  \
\mathcal{H}_{b}(t)=\lan\bm{a}\cdot\bb\ran . \label{3-helicity}
\end{align}
These equations agree with the results in Ref.~\cite{PhysRevD.109.L121302}. Note that although the background rotation $\bO$ does not explicitly enter the expressions of these helicities, it fundamentally influences the dynamics of $\bv$ and $\bb$, as evident in \eqs{spectrum-v}{spectrum-b}. In accordance with \cite{PhysRevD.109.L121302}, the three helicities, as well as the magnetic and kinetic energies can be expressed in $\bk$ space as 
\begin{align}
\mathcal{H}_{c} & = \frac{1}{V}\int _{\bk} (v_{+}b^{*}_{+} + v_{-}b^{*}_{-})  ,   \label{hc-spectrum}  \\
\mathcal{H}_{v} & = \frac{1}{V}\int_{\bk} |\bk|\cdot (|v_{+}|^2 - |v_{-}|^2)   ,    \label{hv-spectrum}   \\
\mathcal{H}_{b} & = \frac{1}{V} \int _{\bk} \frac{1}{|\bk|} \cdot ( |b_{+}|^2- |b_{-}|^2 )   , \label{hb-spectrum}  \\
\mathcal{E}_{b} & =\lan \bb'^2\ran/2 = \frac{1}{2V} \int _{\bk} ( |b'_{+}|^2 + |b'_{-}|^2 )  , \label{Eb-spectrum}   \\
\mathcal{E}_{v} & =\lan \bv^2\ran/2=  \frac{1}{2V} \int _{\bk} ( |v_{+}|^2 + |v_{-}|^2 )  , \label{Ev-spectrum}   
\end{align}
where the last two expressions $\mathcal{E}_{b}$ and $\mathcal{E}_{v}$ are average magnetic and kinetic energies (normalized by $\r$). We also introduce the normalized, dimensionless helicities (denoted with a prime) as follows: $\mathcal{H}'_{c}=\mathcal{H}_{c}/\sqrt{\r}$, $\mathcal{H}'_{v}=\eta\mathcal{H}_{v}$, and $\mathcal{H}'_b=\mathcal{H}_b/(\eta\r)$. Since $\xi_{\o}\propto\m\m_{5}$, we need to solve numerically the coupled Eqs. (\ref{spectrum-v}, \ref{spectrum-b}, \ref{dmu5/dt}). We thus further transform \eq{dmu5/dt} into the explicit evolution equation for $\x'_\o$:
\begin{align}
& \pt_{t}\xi_{\o}^\prime(t) =-\k'_{B} \pt_{t}\mathcal{H}'_{c}(t)-\k'_{\o}\pt_{t}\mathcal{H}'_{v}(t)-C' \pt_{t} \mathcal{H}'_{b}(t)  
  - \G \xi_{\o}^\prime(t),   \label{numerical-4}
\end{align}
where the primed quantities are all normalized to be dimensionless. For the purpose of illustration, we consider a simplified situation where both $v_\pm$ and $b_\pm$ depend only on $k_z$. We perform the numerical simulation by using the following parameter set:  $B_{0}^\prime=5$, and $\xi_{\o}^{\prime}(t=0) =2.0$ and $0.8$, respectively. For the chirality flipping rate $\G$, we consider two cases: $\G=0$ and $\G=0.01/\eta$. The initial field distributions are given by $v_{\pm}(0,k_{z})=b^{\prime}_{\pm}(0,k_{z})= v_{0}/[\mathrm{Exp}(10 \eta |k_{z}|-50 )+1]$ with $v_{0}=0.1$. Other parameters are $k^*_\perp=5/\eta$ (the transverse momentum cutoff), $V=3\eta^3$ (the spatial volume), $\eta^2\sqrt{\r}=10$,  $C^\prime=25/(8\p^2)$, $\k^\prime_{B} = 25/(8\p^2)$, $\k^\prime_{\o}= 1/(4\p^2)$. 

The numerical results are presented in Figs.~\ref{fig-big-ome}-\ref{fig-diff-omega}, illustrating the distinct role of rotation in affecting the CMVI for the case of $\xi'_\omega > 1$ and for the case of $\xi'_\omega < 1$, respectively. For the helicities, we present the ratio relative to the initial value of $\mathcal{H}'_c$. We observe the following properties of the magnetic and kinetic energies, as well as various helicities.

(i) The results for $\x'_\o>1$ are shown in Fig.~\ref{fig-big-ome}. In this case, a CMVI arises even in the absence of rotation, as first observed in Ref.~\cite{PhysRevD.109.L121302}. To make the effect of rotation more transparent, we turn off chirality flipping by taking $\G=0$ (the effect of chirality flipping has been discussed in Ref.~\cite{PhysRevD.109.L121302}). When $\O=0$, the system eventually evolves into a stationary state characterized by $\x'_\o=1$ in which the velocity and magnetic fields are aligned (an Alfvénian state). This can be easily seen from \eqs{spectrum-v}{spectrum-b}: at $\O=0$, when $\x'_\o=1$ and $v_\pm=b'_\pm$, the terms proportional to $\eta$ cancel. In the presence of rotation, the system no longer evolves into a stationary state. This can be verified by requiring $\mathrm{Im}(\o_\c^\l)$ in \eq{Im-CVE} to vanish, which yields no $\bk$-independent solutions for $\x'_\o$. Instead, the system evolves rapidly on the time scale $\sim\eta$ into the region with $\x'_\o\lesssim 1$, and then evolves slowly (quasi-stationarily for small $\O$). The magnetic and kinetic energies as well as various helicities increase rapidly at early times ($t\lesssim\eta$), indicating that the CMVI induces a swift amplification of energies and all three helicities. In sharp contrast to the non-rotating case, where $\mathcal{H}_b$ and $\mathcal{H}_v$ remain zero if they are initially zero, the presence of rotation also leads to a rapid early-time amplification of $\mathcal{H}_b$ and $\mathcal{H}_v$.

(ii) The results for $\x'_\o<1$ are shown in Figs.~\ref{fig-vpo-t} and \ref{fig-diff-omega}. In Fig.~\ref{fig-vpo-t}, we present the momentum profiles of the squared right-handed velocity component, $|v_+(t,k)|^2$ (with $k=k_z$), for different initial values of $\x'_\o$. We turn off chirality flipping by setting $\G=0$, and fix the background magnetic field and rotation to $B'_0=5$ and $\O=5/\eta$, respectively. We focus on $v_+$ because, as shown by the analysis in Sec.~\ref{sec:cve} and Appendix~\ref{spectrum-analysis}, regardless of whether $\x'_\o$ is larger or smaller than $1$, $v_+$ always corresponds to an unstable branch that is amplified by the CMVI. For initial $\x'_\o>1$, we observe that the large-$k$ modes are enhanced most strongly as time increases. By contrast, when $\x'_\o<1$ and decreases further, the unstable region shifts toward the small-$k$ regime. This behavior is consistent with the analysis in Sec.~\ref{sec:cve}, which shows that for $\x'_\o<1$ the instability is confined to the small-$k$ region, as summarized in Tab.~\ref{tab-cve+case12}. Note that $|v_+(t,\bm0)|$ remains constant in time, as can be seen explicitly from \eq{spectrum-v}. 

(iii) In Fig.~\ref{fig-diff-omega}, we show the time evolution of the CVE coefficient, the energies, and various helicities. We choose a small chirality flipping rate, $\G=0.01/\eta$, corresponding to a small mass of the constituent particles. This finite $\G$ causes $\x'_\o$ to decrease slowly even in the absence of rotation. The catalytic effect of rotation on the CMVI is clearly visible in these figures, especially in the case of strong rotation. We also observe oscillatory behavior, most prominently in the cross helicity. This arises from the real part of the dispersion relations, as examined in Appendix~\ref{spectrum-analysis}. The oscillation frequency is determined by a combination of the Alfvén-wave frequency $\o_A$ and the inertial-wave frequency $\o_I$, as demonstrated explicitly in \eqs{ideal-vtkvtk}{ideal-vtkbtk}.

\section{conclusion and outlook}\label{sec:conc}
We have demonstrated that the background rotation plays a vita role in the CMVI and significantly enhances its occurrence. In conventional MHD, rotation $\bO$ splits the usual Alfv\'en wave into fast and slow MC waves with frequencies $\o_{f} > \o_A$ and $\o_s <\o_A$, respectively (see \eqs{fast-MC}{slow-MC}). The emergence of the CMVI requires that the chiral Alfv\'en wave frequency $\o_{CA}$ is larger than either $\o_f$ or $\o_s$ (see \eqs{o+--122}{o---1222}). Since $\o_s$ is always smaller than $\o_A$, the CMVI is thus always easier to occur comparing to the non-rotating case. In fact, we find that there exist unstable modes for any finite $\xi'_\o$ when rotation is present, and the unstable kinematic region is proportional to the rotation frequency $\O$ (see Table \ref{tab-cve+case12}). In contrast, for the non-rotating case, the CMVI appears only when $\x'_\o >1$. Therefore, rotation dramatically catalyzes the CMVI.  

In Figs.~\ref{fig-big-ome}–\ref{fig-diff-omega}, the numerical results illustrate the full evolution of the coupled equations (\ref{spectrum-v}), (\ref{spectrum-b}), and (\ref{numerical-4}). The presence of rotation drastically modifies the system’s evolution, in particular through the emergence of the CMVI for both $\x'_\o>1$ and $\x'_\o<1$ cases. For $\xi_{\o}^{\prime}>1$, unlike the non-rotating case—where the CMVI induces only a rapid growth of the cross helicity when the magnetic and kinetic helicities are initially zero—the rotation-catalyzed CMVI also leads to a rapid generation of magnetic and kinetic helicities. For $\xi_{\o}^{\prime}<1$, rotation opens kinematic windows for the CMVI to occur and such kinematic windows are proportional to the rotation frequency $\O$. In particular, this gives rise to a rapid growth of low-$k$ modes, which in turn leads to a swift enhancement of $\mathcal{H}_{b}$ and $\mathcal{H}_{v}$, in sharp contrast to the non-rotating case studied in Ref.~\cite{PhysRevD.109.L121302}.

Our results suggest a variety of possible implications in different physical systems. First, similar to the CMVI, which aids a dynamo effect, the rotation-catalyzed CMVI could aid a new dynamo effect in a rotating chiral plasma. It would be particularly interesting to explore its influences in electromagnetic plasmas of rotating astrophysical matters and in the rotating quark-gluon plasma in heavy-ion collisions. We leave this as a subject for future investigation.

{\bf Acknowledgments ---} 
We thank Hao-Lei Chen, Yu-Han Gao, Andrey V. Sadofyev, Zhong-Hua Zhang, and Zhi-Bin Zhu for useful discussions. This work is supported by the National Natural Science Foundation of China (Grants No. 12225502 and No. 12147101), the Natural Science Foundation of Shanghai (Grant No. 23JC1400200), and the National Key Research and Development Program of China (Grant No. 2022YFA1604900).

\appendix\section{Maxwell and MHD equations in rotating frame} 
\label{derivofeqns}
In this Appendix, we derive the Maxwell and MHD equations in rotating frame that we have used in the main text. We recall that a rigidly rotating frame with a constant angular velocity $\bO$ is described by the following metric:
\begin{align}
\label{metric}
g_{00}=1-\bu_\O^2,\quad g_{0i}=g_{i0}=-u_\O^i,\quad g_{ij}=-\d_{ij},
\end{align}  
where $\bu_\O=\bO\times\br$ is the frame velocity with respect to the non-rotating inertial frame. The inverse metric is 
\begin{align}
	\label{iimetric}
	g^{00}=1,\quad g^{0i}=g^{i0}=-u_\O^i,\quad g^{ij}=-\d_{ij}+u^i_\O u^j_\O.
\end{align}  
The nonzero components of the Christoffel connection ${\G^\m}_{\n\l}$ are ${\G^i}_{00}=(\bO\times\bu_\O)^i$ and ${\G^i}_{0j}={\G^i}_{j0}=-\e^{ijk}\O^k$. The Maxwell equations in their covariant form are given by $\nabla_\m F^{\m\n}=j^\n$ and $\e^{\m\n\r\s}\nabla_\n F_{\r\s}=0$. Here, $\nabla_\m$ is the covariant derivative such that $\nabla_\m F^{\m\n}=\pt_\m F^{\m\n}+{\G^\m}_{\m\r}F^{\r\n}+{\G^\n}_{\m\r}F^{\m\r}$. Since $|g|=|{\rm det}(g_{\m\n})|=1$, one can actually replace $\nabla_\m$ by $\pt_\m$ to get $\pt_\m F^{\m\n}=j^\n$ and $\e^{\m\n\r\s}\pt_\n F_{\r\s}=0$. Introduce the electric field $\bE$ and magnetic field $\bB$ in rotating frame in the following way:
\begin{align}
\label{ebrot}
E^i = F_{0i},\qquad B^i=-\frac{1}{2}\e^{ijk}F_{jk},
\end{align}   
with $\e^{ijk}$ the 3-dimensional Levi-Civita symbol. In terms of $\bE$ and $\bB$, we can derive 
\begin{align}
\label{Fcompt}
F^{i0}&=g^{i\m}g^{0\n}F_{\m\n} = E^i-(\bu_\O\times\bB)^i \equiv{E^i_\O} ,\\
F^{ij}&=g^{i\m}g^{j\n}F_{\m\n} = u_\O^i E^j_\O-u_\O^j E^i_\O-\e^{ijk}B^k\equiv-\e^{ijk}B_\O^k,
\end{align}   
with $\bB_\O=\bB-\bu_\O\times\bE_\O$. Thus we obtain the Maxwell equations in terms of $\bE$ and $\bB$:
\begin{align}
 \label{eq:mx:2app}
  \bnabla\cdot\bB & =0  ,\\
 \label{eq:mx:3app}
  \bnabla\times \bE & =-\pt_t\bB ,\\
 \label{eq:mx:1app}
 \bnabla\cdot \bE_\O & = n , \\
 \label{eq:mx:4app}
  \bnabla\times\bB_\O & = \pt_t \bE_\O+\bj. 
\end{align}
Substituting $\bE_\O=\bE-\bu_\O\times\bB$ and $\bB_\O=\bB-\bu_\O\times\bE_\O$, the last two Maxwell equations become
\begin{align}
	\label{eq:mx:1}
	\bnabla\cdot \bE & = n + n_{\O} , \\
	\label{eq:mx:4}
	\bnabla\times\bB & = \pt_t \bE+\bj + \bm{i} ,
\end{align}
with two new terms $n_{\O},\bm{i}$ being 
\begin{align}
	&n_{\O}  =\bnabla\cdot(\bu_{\O}\times\bB)= 2\bO\cdot\bB-\bu_{\O} \cdot (\bnabla\times\bB) ,  \\
	&\bm{i} = \bu_{\O} \times (\bnabla\times\bE) + \bnabla\times[\bu_{\O}\times(\bE -\bu_{\O}\times\bB)].
\end{align}
This form of Maxwell equations in rotating frame was first derived in Ref.~\cite{doi:10.1073/pnas.25.7.391} and was then repeatedly re-derived by many others (see, for instance, Ref.~\cite{10.1119/1.1976187}).

To derive the MHD equations in rotating frame, we begin with the relativistically covariant formulation. Let the energy-momentum tensor of the plasma be $T^{\m\n}$. We adopt the Landau-Lifshitz convention for the definition of flow four-velocity $u^\m$ (normalized as $u^2=1$), such that $u^\m$ is the eigenvector of $T^{\m\n}$. This means that $T^{\m}_{\;\;\n} u^\n = \ve u^\m$ with $\ve$ the energy density. With $u^\m$, $T^{\m\n}$ can be decomposed into the following form,
\begin{align}
	\label{tmndecot}
	T^{\m\n}=(\ve+p)u^\m u^\n-p g^{\m\n}+\rm{viscous\;\; terms},
\end{align}   
where $p$ is the thermodynamic pressure and we have omitted the viscous terms which is not our focus in this article. The hydrodynamic equations are derived from the energy-momentum conservation law, 
\begin{align}
	\label{tmnc}
\nabla_\m T^{\m\n}=F^{\n\m}j_\m.
\end{align}   
By contracting with $u_\n$ and $\D^\n_\l=\d^\n_\l-u^\n u_\l$, this equation leads to (viscous terms are omitted)
\begin{align}
	\label{tmnc21}
	&u^\m\pt_\m\ve+(\ve+p)\nabla_\m u^\m=F^{\n\m}u_\n j_\m,\\
		\label{tmnc22}
	&(\ve+p)u^\m \nabla_\m u^\n-\D^{\n\m}\pt_\m p = \D^{\n\l}F_{\l\m}j^\m.
\end{align}   

To proceed, we consider the non-relativistic limit, $u^\m\approx (1,\bv)$ with $|\bv|\sim u_0\ll 1$ and $|\bu_\O|\sim u_0\ll 1$, where $u_0=L/\t$ is a typical velocity scale of the system. This requires either a small $|\bO|$ or a small region close to the rotating axis. Direct calculations give the following results:
\begin{align}
\label{umusimpl}
	& u^\m\nabla_\m \bv=\pt_0\bv+\bv\cdot\bm\nabla\bv+2\bO\times\bv+\bO\times\bu_\O +O(u_0^3),\\
		&\nabla_\m u^\m =\bm\nabla\cdot\bv +\frac{1}{2}\pt_0(\bu_\O+\bv)^2+O(u_0^3).
\end{align}  
Furthermore, \eq{eq:mx:3app} leads to $|\bE|/|\bB|\sim L/\t=u_0$, which in turn leads to $|\bE_\O|/|\bB|\sim u_0$ and $|\bB_\O|/|\bB|\sim 1+O(u_0^2)$. In addition, 
\begin{align}
	& |\frac{\pt_{t}\bE}{\bnabla\times\bB} | \sim u_{0}^2 , \\
	& |\frac{\bm{i}}{\bnabla\times\bB}| \sim  
	|\bu_{\O}|u_{0}+|\bu_{\O}|^2   \sim u_0^2.
\end{align}
Thus, combining these results, we have $\bnabla\times\bB\approx\bj$ in the non-relativistic limit.

Define the contravariant electric and magnetic fields:
\begin{align}
	\label{newenn}
\tilde{E}^\m=F^{\m\n}u_\n,\quad \tilde{B}^\m=\frac{1}{2}\e^{\m\n\r\s}u_\n F_{\r\s}.
\end{align}
It is straightforward to show that
\begin{align}
	\label{newennpowercounpp}
&\tilde{E}^0\approx (\bE+\bv\times\bB)\cdot(\bv+\bu_\O)\sim O(u_0^2),\\
&\tilde{\bE}=\bE+\bv\times\bB +O(u_0^3),\\
&\tilde{B}^0\approx\bB\cdot(\bv+\bu_\O)+O(u_0^3),\\
\label{newbbn}
&\tilde{\bB}=\bB +O(u_0^2).
\end{align}
Substituting \eqs{umusimpl}{newbbn} into \eqs{tmnc21}{tmnc22}, we obtain (accurate up to $O(u_0^3)$)
\begin{align}
	\label{hydro1}
&\ls 1+\frac{(\bu_\O+\bv)^2}{2}\rs\pt_0\ve +\bv\cdot\bm\nabla\ve+(\ve+p)\bm\nabla\cdot\bv\nonumber\\&\qquad\qquad\qquad\,+\frac{1}{2}(\ve+p)\pt_0(\bu_\O+\bv)^2=\bE\cdot\bj, \\
\label{hydro2}
&(\ve+p)(\pt_0\bv+\bv\cdot\bm\nabla\bv)+2\bO\times\bv+\bO\times\bu_\O+\bm\nabla p\nonumber\\&\qquad\qquad\qquad\qquad\qquad\qquad\qquad\qquad\;\,=\bj\times\bB.
\end{align}
At non-relativistic limit, $\ve=\rho +\ve'$ where $\rho$ is the mass density and $\ve'$ is the internal kinetic energy. They satisfy $\ve'\sim p \sim O(u_0^2)\ll \rho$. Usually, the magnetic energy $\sim\bB^2$ is smaller than $\r$, so that we treat $\bj\cdot\bE$ as subleading order. Keeping only leading order terms in \eq{hydro1}, we obtain,
\begin{align}
	\label{hydro11}
	&\pt_0\rho +\bm\nabla\cdot(\r\bv)=0,
\end{align} 
which is the continuity equation. For incompressible fluid, $\r$ is constant under fluid flow, this equations gives $\bm\nabla\cdot\bv=0$, and vice verse. Now, approximating $\ve+p$ by $\r$ and substituting $\bj\approx\bnabla\times\bB$ into \eq{hydro2}, we obtain \eq{NS-rot-inEMF} in the main text. 


We now derive the expression for $\bj$. The contravariant electric current is 
\begin{align}
	\label{current1110}
j^\m=n u^\m +\s \tilde{E}^{\m}+\xi_B\tilde{B}^\m+\xi_\o\tilde{\o}^\m,
\end{align} 
where $\tilde{\o}^\m=\e^{\m\n\r\s}u_\n\nabla_\r u_\s$ and $\s=1/\eta$ is the electric conductivity. Substituting \eqs{newennpowercounpp}{newbbn}, we find that 
\begin{align}
	\label{current111}
\bj&\approx n \bv +\s (\bE+\bv\times\bB)+\xi_B\bB+\xi_\o(\bo+2\bO)\nonumber\\
&\approx \s (\bE+\bv\times\bB)+\xi_B\bB+\xi_\o(\bo+2\bO),
\end{align}
where we have used the fact that $n\bv\sim O(u_0^2)$ due to \eq{eq:mx:1app}. This gives that 
\begin{align}
	\label{current11ddd1}
	\bE&\approx -\bv\times\bB+\eta[\bj-\xi_B\bB-\xi_\o(\bo+2\bO)].
\end{align}
Substituting this into \eq{eq:mx:3app}, one obtains \eq{pt-B-chiral} in the main text. 

Analogous to \eq{current1110}, the contravariant chiral (axial) current is   
\begin{align}
	j^{\m}_{5}=n_{5}u^{\m}+\k_{B}\tilde{B}^{\m}+\k_{\o} \tilde{\o}^{\m},
\end{align} 
where $n_{5}$ is the chiral charge density of constituent particles, $\k_{B}$ is the coefficient of the chiral separation effect, and $\k_{\o}$ characterizes the  axial CVE. Substituting \eqs{newennpowercounpp}{newbbn}, we find that $j^{\m}_{5}$ can be approximated as
\begin{align}
	j^{0}_{5} & \approx  n_{5} + \k_{B} \bB\cdot(\bv+\bu_{\O}) + \k_{\o} ( \bo + 2 \bO)\cdot (\bv + \bu_{\O} ),    \\
	\bj_{5} & \approx n_{5}\bv + \k_{B} \bB + \k_{\o}(\bo+2\bO).
\end{align}
The chiral anomaly equation reads 
\begin{align}
	\nabla_{\m}j^{\m}_{5}=\pt_{\m}j^{\m}_{5}=-C \tilde{E}^\m \tilde{B}_{\m},   \label{chiral-anom}
\end{align}
where $C=e^2/2\pi^2$ is the anomaly coefficient.  Substituting \eqs{newennpowercounpp}{newbbn}, it becomes \eq{chiral-anom-3D} in the main text after neglecting terms $\sim O(u_0^3)$.

\section{Sign of the imaginary part of the frequencies}\label{appenanlay}
We consider the pure CME case first. Let us denote the two solutions to  Eq.~(\ref{chiral-dispersion-cmetotal}) as $\o_{1}$ and $\o_{2}$ with $\o_{1,2}=\o_{1,2a}+ i\,\o_{1,2b}$ ($\o_{1,2a/b}\in \mathbb{R}$). Thus Eq.~(\ref{chiral-dispersion-cmetotal}) is factorized into $(\o-\o_{1})(\o-\o_{2})=0$ which  immediately gives us the following relations ($\x_B>0$),
\begin{align}
	\label{cme-o12-100} 
	& \o_{1a}+\o_{2a}=\l\o_I,\\
	&\o_{1b}+\o_{2b}= 	- \eta k(k+\l\xi_{B}) ,    \label{cme-o12-1} \\
	&    \o_{1a}\o_{2a}- \o_{1b}\o_{2b}=-\o_A^2, \label{cme-o12-1333}\\
	& \o_{1a}\o_{2b}+\o_{1b}\o_{2a}=-\eta k(\xi_{B}+\l k) \o_{I}. 
\end{align}
Note that $(\o_{1a}+\o_{2a})(\o_{1b}+\o_{2b})=\o_{1a}\o_{2b}+\o_{1b}\o_{2a}$ which gives $\o_{1a}\o_{1b}+\o_{2a}\o_{2b}=0$. Using this relation and multiplying Eq.~(\ref{cme-o12-1333}) by $\o_{1b}$, we obtain $(\o_{2a}^2+\o_{1b}^2)\o_{2b}=\o_A^2\o_{1b}$, which shows that $\o_{1b}$ and $\o_{2b}$ must have the same sign. These equations reveal the properties of both real and imaginary parts, allowing us to directly assess the behavior of wave propagation or instability. For instance, in Eq. (\ref{cme-o12-1}), when $k+\l\xi_{B}<0$ (this is possible only for $\l=-1$), the two solutions have positive imaginary parts, $\o_{1b}>0$ and $\o_{2b}>0$. Therefore, the CME induces an instability in the region $k\in(0,\xi_{B} )$. Additionally, when $k+\l \xi_{B}=0$ (again, this is possible only for $\l=-1$), these equations yield $\o_{1b}=\o_{2b}=0$, and thus all waves propagate without damping or growth. Furthermore, when $k+\l\xi_{B}>0$, Eq. (\ref{cme-o12-1}) implies both $\o_{1b}$ and $\o_{2b}$ should be negative (because they have the same sign), and thus no instability occurs. Thererfore, the CPI occurs for $\l=-1, 0<k< \xi_{B}$ which is the same as the $\bO=\bm 0$ case.

We turn to the CVE case. In Sec.~\ref{sec:cve}, we have provided a detailed analysis of CMVI in the rotating frame based on the solution of Eq. (\ref{disper-cve}). Here, we perform another analysis by adopting a geometric approach to Eq. (\ref{disper-cve}). First, we express $\o=\o_{a}+i\cdot\o_{b}$, $\o_{a,b}\in\mathbb{R}$, and split Eq. (\ref{disper-cve}) into two distinct curves in $(\o_a, \o_b)$ plane:
\begin{align}
	\mathrm{C_{1}} : \quad 
	& (\o_{a}- \frac{\l\o_{I}}{2})^2- (\o_{b}+ \frac{\g_{\eta}}{2})^2 
	= \frac{\o_{I}^2-\g_{\eta}^2}{4}+\o_{A}^2 ,    \label{curve1} \\
	\mathrm{C_{2}}: \quad  &  
	\o_{b}= \frac{ (\l\o_{I}/2-\o_{CA}) \g_{\eta} }{2\o_{a}-\l\o_{I}}-\frac{ \g_{\eta} }{2} ,\label{curve2}
\end{align}
where $\mathrm{C_{1,2}}$ are two hyperbolas. They have zero points at 
\begin{align}
	\o_{\mathrm{C_{1}}\pm} & = \l \frac{\o_{I}}{2}
	\pm  \sqrt{ \frac{\o_{I}^2}{4} + \o_{A}^2 }   ,
	\label{c1-zero-point}   \\
	\o_{\mathrm{C_{2}}} & = \l \o_{I}- \o_{CA}.
	\label{c2-zero-point}
\end{align}
These zero points vary with $k$, causing the two curves to shift in position. The intersection points of the two curves correspond to the solutions of $\o$, which determine whether the solutions represent stable or unstable modes. By comparing the values of $\o_{\mathrm{C_{1}}\pm}, \o_{\mathrm{C_{2}}}$, we can determine whether instability arises. By categorizing the four main cases based on $\l, \cos\h$ (with $\h$ the angle between $\bk$ and $\bB_0$), we can systematically analyze the instabilities in different parameter regimes in Eqs. (\ref{curve1}-\ref{c2-zero-point}).

(1) {\bf Case 1:} $\l=+1$ and $\cos\h>0$. In this case, $\o_A, \o_{CA}, \o_I >0$. When $0<\o_{CA} \leq 2\O$, two intersection points of the two curves have $\mathrm{Im}(\o)<0$, indicating that both modes are damped. When $\o_{CA}>2\O$, the value of $\o_{\mathrm{C_{2}}}$ becomes negative, $\o_{\mathrm{C_{2}}}<0$. One of the intersection points still has $\mathrm{Im}(\o)<0$, corresponding to a damped mode. For the other intersection point, the stability depends on the values of $\o_{\mathrm{C_{1}}-}$ and $\o_{\mathrm{C_{2}}}$. If $\o_{\mathrm{C_{1}}-} \leq \o_{\mathrm{C_{2}}}<0$, one obtains a damped mode. This condition leads to
\begin{equation}
	0<\o_{CA}- \frac{\o_{I}}{2} \leq \sqrt{ \frac{\o_{I}^2}{4}+\o_{A}^2 },    \label{2Oxi>sth}
\end{equation}
which is satisfied when
\begin{align}
	& \xi_{\o}^{\prime 2} \leq 1 \quad \&  \quad
	\frac{2\O}{B_{0}^{\prime}\xi_{\o}^{\prime}}< k\,, \\
	\mathrm{or}  \quad  &   \xi_{\o}^{\prime 2}>1   \quad   \&  \quad \frac{2\O}{B_{0}^{\prime}\xi_{\o}^{\prime}} < k \leq  \frac{2\O\xi_{\o}}{B_{0}(\xi_{\o}^{\prime2}-1) } .
\end{align}
On the other hand, if $\o_{\mathrm{C_{2}}}<\o_{\mathrm{C_{1}}-} $, the other intersection point has $\mathrm{Im}(\o)>0$, leading to an instability. This condition requires
\begin{equation}
	\frac{\o_{I}}{2} + \sqrt{ \frac{\o_{I}^2}{4} + \o_{A}^2 } <  \o_{CA}, \label{case1-cve+O+instab-0}
\end{equation}
which is satisfied when
\begin{equation}
	\xi_{\o}^{\prime 2}> 1 \quad \& \quad 
	\frac{2\O\xi_{\o} }{B_{0}(\xi_{\o}^{\prime 2}-1)} <  k .
	\label{case1-cve+O+instab}
\end{equation} 
Hence, the instability is primarily observed in Eqs. (\ref{case1-cve+O+instab-0})-(\ref{case1-cve+O+instab}), where unstable modes are excited in the large $k$ range with $\xi_{\o}^{\prime2}>1$. This behavior aligns with the CMVI discussed in \cite{PhysRevD.109.L121302}, and a non-zero $\bO$ raises the lower threshold for $k$.
\begin{figure*}[t!]
	\centering
	\includegraphics[width=.40\textwidth]{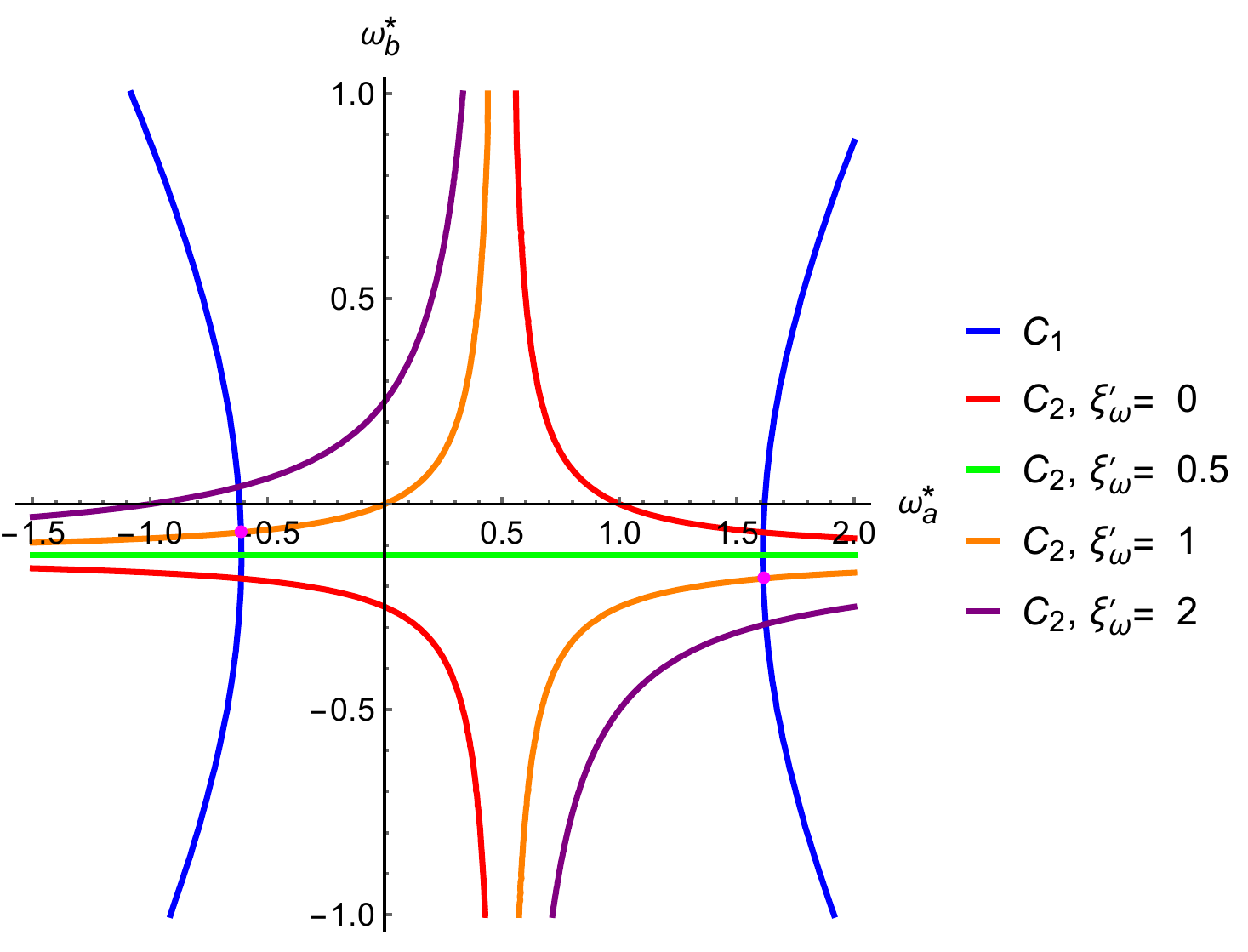}\qquad\qquad
	\includegraphics[width=.40\textwidth]{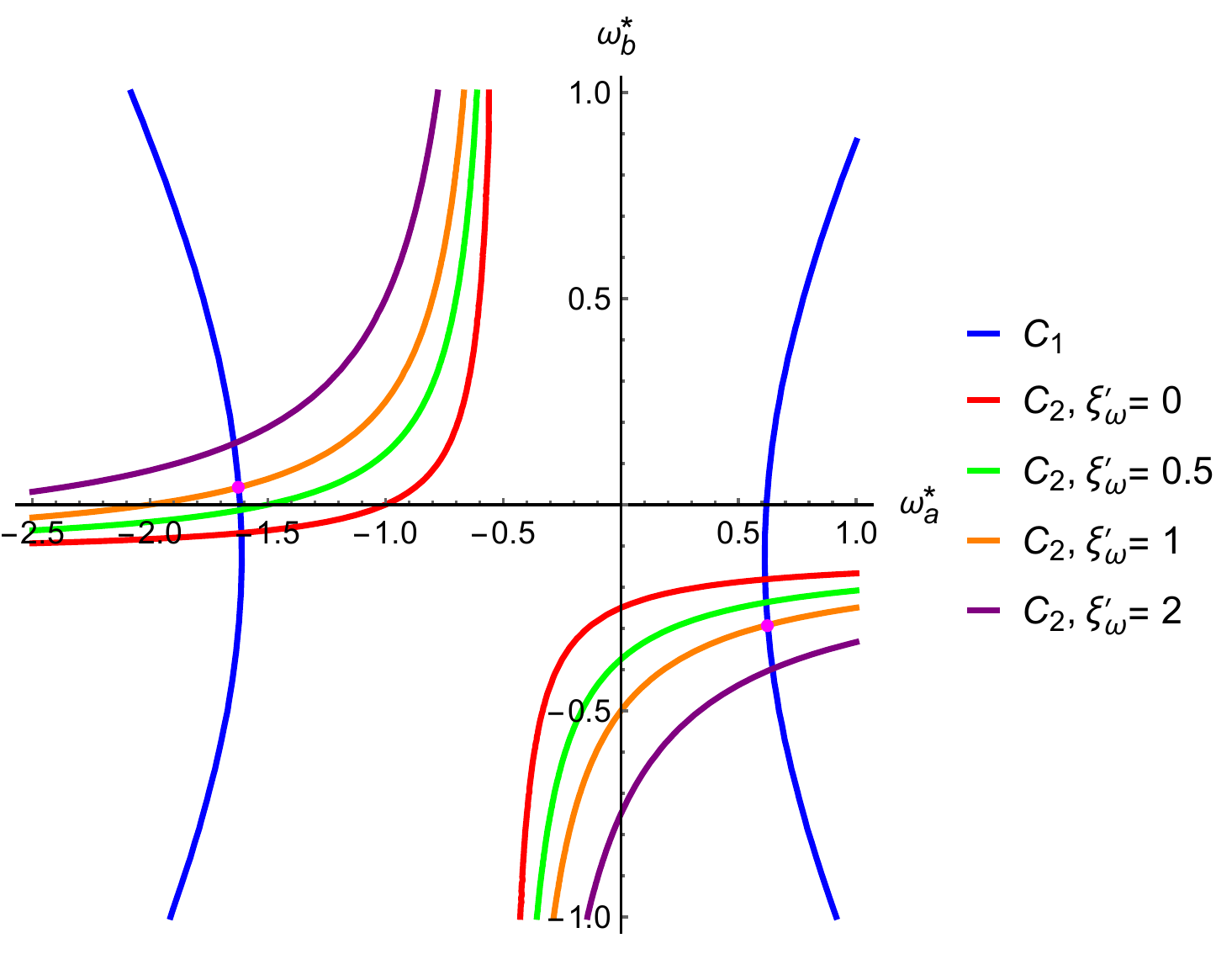}
	
	\caption{
		\textbf{Left:} Curves $\mathrm{C}_{1,2}$ in the $(\o^{*}_{a}, \o^{*}_{b})$ plane for Case 1, where $\o^{*}_{a,b}\equiv \eta\,\o_{a,b}$. The curves $\mathrm{C}_{1}$ and $\mathrm{C}_{2}$ have two intersection points that shift with $\xi_{\o}^{\prime}$. When $\xi_{\o}^{\prime}\le 1$, both the two intersection points have negative $\o_{b}^{*}$, corresponding to stable damped modes. Only when the zero point condition $\o_{C_{2}}< \o_{C_{1}-}$ is satisfied and $\xi_{\o}^{\prime}>1$, one of intersection points becomes positive for $\o_{b}^{*}$, indicating instability. \textbf{Right:}  Curves $\mathrm{C}{1,2}$ in the $(\o^{*}_{a},\o^{*}_{b})$ plane for Case 3. One of the intersection points become positive for $\o_{b}^{*}$ when the zero point condition $\o_{C_{2}}< \o_{C_{1}-}$ is satisfied, which allows for $\xi_{\o}^{\prime}\le1$. The parameters are $ \eta k= 0.5, \eta \O=1, B_{0}^{\prime} =4,\h=\pi/3$.  } 
	\label{case1+3}
\end{figure*}

For clarity, we plot the two curves, $\mathrm{C}_{1,2}$, for Case 1 on the $\omega_a$-$\omega_b$ plane in Fig.~\ref{case1+3} (left), where the frequency is normalized by $\eta$ such that $\omega_{a,b}^* = \eta\,\omega_{a,b}$. The solutions to the coupled equations (\ref{curve1}) and (\ref{curve2}) correspond to the two intersection points of curves $\mathrm{C}_1$ and $\mathrm{C}_2$. As $\xi_{\omega}^{\prime}$ increases, the two intersection points move along curve $\mathrm{C}_1$ in the following manner: One intersection point remains with $\omega_b^* < 0$, corresponding to the stable damping mode, while the other undergoes a sign change of $\omega_b^*$ when the condition (\ref{case1-cve+O+instab}) is satisfied, as indicated by the intersection point of the purple and blue curves.

(2) {\bf Case 2:} $\l=+1$ and $\cos\h<0$. In this case, $\o_A, \o_{CA}, \o_I <0$. We have the same results as Case 1, the only difference lies in the exchange of positions between the two intersection points. 

(3) {\bf Case 3:} $\l=-1$ and $\cos\h >0$. In this case, $\o_A, \o_{CA}, \o_I >0$ and $\l\o_{I}/2-\o_{CA}$ is always negative. Additionally, the zero point $\o_{\mathrm{C_{2}}}<0$, and the condition $\o_{\mathrm{C_{1}}-}, \o_{\mathrm{C_{2}}}<-\o_{I}$ holds. It is evident that one of the intersection points has $\mathrm{Im}(\o)<0$, representing a damped mode. For the other one, the stability depends on the values of $\o_{\mathrm{C_{1}}-},\o_{\mathrm{C_{2}}}$. If $\o_{\mathrm{C_{1}}-} \leq \o_{\mathrm{C_{2}}}$, it has $\mathrm{Im}(\o)\leq0$, representing a damped mode. This condition leads to
\begin{equation}
	\frac{\o_{I}}{2}+\o_{CA}   \leq 
	\sqrt{\frac{\o_{I}^2}{4}+\o_{A}^2},     
\end{equation}
which means 
\begin{equation}
	\xi_{\o}^{\prime 2} <1 \quad \&  \quad 
	\frac{ 2\O  \xi_{\o} }{ B_{0} (1- \xi_{\o}^{\prime2})}  \leq k . \label{case3+cve+stable}
\end{equation}
On the other hand, if $\o_{\mathrm{C_{2}}}<\o_{\mathrm{C_{1}}-} $, it has $\mathrm{Im}(\o)>0$, representing an unstable mode. This leads to 
\begin{align}
	&  \sqrt{ \frac{\o_{I}^2}{4}+ \o_{A}^2}  <
	\frac{\o_{I}}{2}+\o_{CA},
	\label{case3-lmd=-1,cve-unstable}
\end{align}
which implies  
\begin{align}
	&  \xi_{\o}^{\prime2} \geq 1 \quad  \& \quad  0 < k  , 
	\label{case3-lmd=-1,cve,instability-0}  \\
	\mathrm{or} \quad  &  \xi_{\o}^{\prime 2}  <1  \quad  \& \quad 
	0<k<\frac{2\O \xi_{\o}}{ B_{0}(1-\xi_{\o}^{\prime2})}.
	\label{case3-lmd=-1,cve,instability}
\end{align}

For demonstration, we plot the two curves, $\mathrm{C}_{1,2}$, for Case 3 on the $\omega_a$-$\omega_b$ plane in Fig.~\ref{case1+3} (right). In this case, one of the two intersection points maintains $\omega_b^* < 0$ for any $\xi_{\omega}^{\prime} > 0$. For the other intersection point, we observe that $\omega_b^* > 0$ when $\xi_{\omega}^{\prime} = 1$, suggesting that $\omega_b^*$ could become positive for $\xi_{\omega}^{\prime} \in (0, 1)$, thereby leading to instability.

(4) {\bf Case 4:} $\l=-1$ and $\cos\h<0$. In this case, $\o_A, \o_{CA}, \o_I <0$. We will have same results as shown in Case 3, the only difference lies in the exchange of positions between the two intersection points.

The results presented in Case 1 to Case 4 precisely coincide the results we present in the main text.

\section{The helicity analysis}\label{helicity-analysis}

We analyze the contributions of the background rotation $\bO$ and magnetic field $\bB_{0}$ to various helicities. We assume that $\bO\pll\bB_{0}$ and that the fluctuation fields $\bb, \bv\perp \bB_0$. The cross helicity, kinetic helicity, and magnetic helicity are given by
\begin{align}
\mathcal{H}_{c} & = \lan(\bv+\bu_{\O})\cdot (\bB_{0}+\bb)\ran = \lan \bv \cdot \bb\ran + \lan \bu_{\O} \cdot \bb \ran,  \label{app-hc}  \\
\mathcal{H}_{v} & =\lan (\bv+\bu_{\O})\cdot (\bo+2\bO) \ran = \lan \bv \cdot \bo \ran + \lan \bu_{\O} \cdot \bo \ran, \label{app-hv} \\
\mathcal{H}_{b} & =\lan (\bm{A}_{0}+\bm{a})\cdot (\bB_{0} + \bb) \ran  \nonumber  \\
& = \lan \bm{a}\cdot \bb \ran + \lan \bm{a}\cdot \bB_{0} \ran  + \lan \bm{A}_{0} \cdot \bb \ran\nonumber\\
&=\lan \bm{a}\cdot \bb \ran  + 2 \lan \bm{A}_{0} \cdot \bb \ran,  \label{app-hb} 
\end{align}
where $\lan\cdots\ran=V^{-1}\int d^3\bx(\cdots)$,  $\bm{A}_0=(1/2)\bB_0\times\bx$, and we have used 
\begin{align}
\lan \bm{a}\cdot \bB_{0} \ran =\lan \bm{A}_{0} \cdot \bb \ran, \label{aB0}
\end{align}
neglecting a surface term. 

We now evaluate the terms involving background fields one by one. It is straightforward to derive
\begin{align}
&\lan \bu_{\O} \cdot \bb \ran  = \frac{1}{V}\int d^3\bx (\bO\times\bx)\cdot \bb(t,\bx)   \nonumber  \\&= \frac{i}{V}\bO\cdot[\bnabla_{\bk} \times \bb(t,\bk)]|_{\bk=0}.    \label{UOb}
\end{align}
Similarly, we have
\begin{align}
& \lan \bu_{\O}\cdot \bo\ran= \frac{i}{V}\bO\cdot [\bnabla_{\bk}\times\bo(t,\bk)]|_{\bk=0}\label{UOo}\, , \\
& \lan \bm{A}_{0}\cdot\bb\ran=\frac{i}{2V}\bB_{0} \cdot [\bnabla_{\bk}\times \bb(t,\bk)]|_{\bk=0}   \label{Ab}\,.
\end{align}
Using the Fourier-space forms of Eqs. (\ref{linear-v}) and (\ref{lin-b-chiral})
\begin{align}
  \pt_{t}\bb(t,\bk) & = [(\bB_{0}\cdot i\bk)-\eta\xi_{\o}(i\bk)^2]\bv+\eta (i\bk)^2 \bb,  \\
  \pt_{t}\bo(t,\bk) & = 2 (\bO\cdot i \bk) \bv + \frac{1}{\r}(\bB_{0}\cdot i\bk) i\bk \times \bb,
\end{align}
we immediately obtain
\begin{align}
\pt_{t} [\bnabla_{\bk} \times \bb(t,\bk)]|_{\bk=0}    & =  i\bB_{0} \times \bv(t,\bk=0) \,, \\
\pt_{t}[\bnabla_{\bk}\times \bo(t,\bk)]|_{\bk=0} & = 2i \bO\times \bv(t,\bk=0) \,.
\end{align}
Integrating over time gives
\begin{align}
 [\bnabla_{\bk} \times \bb(t,\bk)]|_{\bk=0} & = \int_{0}^t dt^\prime \ i\bB_{0} \times\bv (t^\prime,0) +\mathrm{\bm{C}}_{b}, \\  [\bnabla_{\bk}\times \bo(t,\bk)]|_{\bk=0} & = \int_{0}^{t} dt^\prime \ 2i \bO\times \bv(t^\prime,0) + \mathrm{\bm{C}}_{\o},
\end{align}
where $\mathrm{\bm{C}}_{b}=[\bnabla_{\bk} \times \bb(t=0,\bk)]|_{\bk=0}$ and  $\mathrm{\bm{C}}_{\o}= [\bnabla_{\bk} \times \bo(t=0,\bk)]|_{\bk=0} $ are constant vectors determined by initial conditions. Therefore, we have
\begin{align}
\lan \bu_{\O}\cdot\bb\ran & = \frac{i}{V} \bO\cdot \mathrm{\bm{C}}_{b}  ,  \\
\lan \bu_{\O}\cdot\bo\ran & = \frac{i}{V} \bO\cdot \mathrm{\bm{C}}_{\o} , \\
\lan \bm{A}_{0}\cdot\bb\ran & = \frac{i}{2V} \bB_{0}\cdot \mathrm{\bm{C}}_{b}, 
\end{align}
which are time independent. So we only need to consider the initial conditions for $\mathrm{\bm{C}}_{b,\o}$. In our numerical simulation, we have chosen the initial velocity $\bv(t=0,\bk)$ and magnetic field $\bb(t=0,\bk)$ to depend only on $k_z$, while $\bO$ and $\bB_0$ are along $z$-axis. Thus, we have 
\begin{align}
 \bO\cdot\mathrm{\bm{C}}_{b} & = \e_{ijl}\O_i\pt_{k_j} b_l(t=0,\bk)|_{\bk=0}=0.
\end{align}
Similarly, $\bB_0\cdot\mathrm{\bm{C}}_{b}=0$. For $\bO\cdot\mathrm{\bm{C}}_{\o}$, we have
\begin{align}
 \bO\cdot\mathrm{\bm{C}}_{\o} & = -i\bO\cdot\bv(t=0, \bk)|_{\bk=0}=0
\end{align}
as $\bv\perp\bO$. Therefore, we have, in our setup, 
\begin{align}
\lan \bu_{\O}\cdot\bb\ran =
\lan \bu_{\O}\cdot\bo\ran =
\lan \bm{A}_{0}\cdot\bb\ran =0,  \label{back-heli-00}
\end{align}
meaning that the background rotation and magnetic field do not contribute to various helicities in the setup of our calculations. 

From the above analysis, it is clear that we only need to focus on $\lan\bv\cdot\bb\ran, \lan\bv\cdot \bo\ran$, and $\lan \bm{a}\cdot\bb\ran$ in our analysis. The full calculation of them in momentum space is presented in Sec.~\ref{sec:evolu}, and it is instructive to examine them directly in coordinate space here. Starting from the evolution equation for the vector potential $\bm{A}(t,\bx)$ with $\bnabla\times \bm{A}=\bB$, we uncurl the Eqs. (\ref{pt-B-chiral}-\ref{lin-b-chiral}) to produce
\begin{align}
\pt_{t}\bm{A} & = -\eta \bnabla\times \bB + \bv \times\bB+\eta\xi_{\o}(\bo + 2\bO) \label{dtA} , \\   
\pt_{t}\bm{a} & = -\eta \bnabla\times\bb + \bv \times (\bB_{0} +\bb) + \eta \xi_{\o}(\bo+2\bO) ,\label{dta}
\end{align}
up to total gradient terms, where $\bm{A}=\bm{A}_{0}+\bm{a},\bm{A}_{0}=\bB_{0} \times\bx/2$. One can see that \eq{dta} is the same as \eq{current11ddd1} after taking $-\pt_{t}\bm{a}= \bE$.  Additionally, taking the curl of \eq{NS-rot-inEMF}, we obtain 
\begin{align}
\pt_{t} \bo = \bnabla \times [\bv \times (\bo+2\bO)]   + \frac{1}{\r} \bnabla \times (\bj \times \bB) . \label{dtomega}  
\end{align}
The time derivatives of the three helicities can then be calculated by using Eqs.~(\ref{NS-rot-inEMF}, \ref{linear-v},\ref{pt-B-chiral}-\ref{lin-b-chiral}, \ref{dtA}-\ref{dtomega}). The resultant expressions are: 
\begin{align}
 &\pt_{t} \lan \bv \cdot \bb \ran = -\eta \lan \bj \cdot \bo \ran +   \bB_{0} \cdot \lan \bo \times \bv\ran + \eta\xi_{B} \lan \bb \cdot \bo\ran  \nonumber \\ 
 & +\eta \xi_{\o} \lan \bo ^2 \ran + 2 \bO \cdot \lan \bb \times \bv \ran  + \frac{ \bB_{0} \cdot \lan\bb\times \bj\ran }{\r},  \label{app-dt-dHc}
\end{align}
\begin{align}
 \pt_{t}\lan \bm{a}\cdot\bb \ran &= - 2\eta \lan \bb \cdot \bj\ran + 2\bB_{0} \cdot \lan  \bb \times \bv  \ran + 2 \eta \xi_{\o}\lan \bb \cdot \bo\ran \nonumber   \\
& + \eta \xi_{\o} 2\bO \cdot \lan \bb\ran +2 \eta \xi_{B} \lan \bb^2 \ran  + \eta \xi_{B} \bB_{0} \cdot \lan \bb \ran,    \label{app-dt-dHb}
\end{align}
\begin{align}
\pt_{t}\lan \bv \cdot \bo\ran & = 4 \bO \cdot \lan \bo \times \bv\ran +\frac{2 \bB_{0} \cdot \lan \bo \times \bj\ran }{\r} +\frac{2\lan \bo \cdot (\bj\times\bb)\ran}{\r}.   \label{app-dt-dHv}
\end{align}
Similarly, the time derivatives of the magnetic energy, kinetic energy, as well as their sum are given by 
\begin{align}
 \pt_{t}\lan \bb^2/2 \ran  & = - \eta \lan \bj ^2 \ran +  \bB_{0} \cdot \lan  \bj \times \bv \ran + \lan  \bv \cdot (\bb \times \bj)\ran  \nonumber \\
&\quad + \eta \xi_{B} \lan \bb \cdot \bj\ran +  \eta \xi_{\o} \lan \bo \cdot \bj \ran,  \label{app-dt-dEb}
\end{align}
\begin{align}
\pt_{t}\lan \bv^2/2\ran  = \frac{1}{\r} \bB_{0} \cdot \lan \bv \times \bj \ran + \frac{1}{\r} \lan \bv \cdot (\bj\times\bb)\ran,  \label{app-dt-dEv}
\end{align}
\begin{align}
& \pt_{t}\lan\bb^2/2\ran+\r\pt_{t}\lan \bv^2/2\ran             \nonumber       \\
&\qquad\qquad = - \eta \lan \bj ^2 \ran  +  \eta \xi_{B} \lan \bb \cdot \bj\ran +  \eta \xi_{\o} \lan \bo \cdot \bj \ran.
\end{align}
These results are consistent with the conventional forms~\cite{SHEBALIN_2006,10.1063/1.4824009, Brandenburg:2020vwp} upon setting $\xi_{B} = \xi_{\omega} = 0$. The above expressions indicate that background rotation and magnetic fields modify the transfer of both energy and helicity. In Eqs.~(\ref{app-dt-dHc}, \ref{app-dt-dHb}), the anomalous terms $\eta \xi_{\omega} \langle \boldsymbol{\omega}^{2} \rangle$ and $2 \eta \xi_{B} \langle \boldsymbol{b}^{2} \rangle$ are positive for $\xi_{\omega} > 0$ and $\xi_{B} > 0$, demonstrating that the presence of the CVE (CME) generates cross (magnetic) helicity, as previously clarified in discussions of the CPI and CMVI~\cite{PhysRevLett.79.1193, PhysRevLett.111.052002, PhysRevD.109.L121302}. By contrast, the terms $-\eta \langle \boldsymbol{j} \cdot \boldsymbol{\omega} \rangle$ and $-\eta \langle \boldsymbol{b} \cdot \boldsymbol{j} \rangle$ are not sign-definite and are difficult to analyze directly in coordinate space. Therefore, we employ a Fourier-transform approach, as discussed in Sec.~\ref{sec:evolu}, to carry out a more systematic analysis.

\begin{widetext}    

\section{The analytical treatment of the evolution equations}\label{spectrum-analysis}
We apply spectral analysis to study the properties of the evolution equations (\ref{spectrum-v}) and (\ref{spectrum-b}) analytically for constant $\x_\o$. First, we rewirte Eqs. (\ref{spectrum-v}) and (\ref{spectrum-b}) as
\begin{align}
\pt_{t}^2  v_{\pm} & = (\pm i\o_{I}-\eta\bk^2) \pt_{t}v_{\pm} 
+ [ i \eta \bk^2 (\o_{A} \xi_{\o}^{\prime} \pm \o_{I}) -\o_{A}^2 ]  v_{\pm},   \label{2nd-ode-v}    \\
\pt_{t}^2 b^\prime_{\pm} & =  (\pm i\o_{I} -\eta \bk^2)\pt_{t} b^\prime_{\pm} + \pt_{t} \xi^\prime_{\o}(t)\cdot  \eta \bk^2 v_{\pm}  + [ i \eta \bk^2 (\o_{A} \xi_{\o}^\prime \pm \o_{I}) 
-\o_{A}^2 ] b^\prime_{\pm} .  \label{2nd-ode-b}
\end{align}
These are second-order ordinary differential equations when $\xi_{\o} =\mathrm{const}$. From Eqs. (\ref{spectrum-v}) and (\ref{spectrum-b}), we also note that $\pt_{t}v_{\pm}(t,0)=\pm i \o_{I} \cdot v_{\pm}(t,0) $ and $\pt_{t}b^\prime_{\pm}(t,0)=0$ when $\bk=0$.

Before delving into \eqs{2nd-ode-v}{2nd-ode-b}, it is useful to consider the ideal MHD case ($\eta=0$): 
\begin{align}
\pt_{t}^2 v_{\pm} & - (\pm i\o_{I}) \pt_{t}v_{\pm} + \o_{A}^2   v_{\pm} =0 ,   \label{ideal-ode-v}    \\
\pt_{t}^2 b^\prime_{\pm} & - (\pm i\o_{I} )\pt_{t} b^\prime_{\pm} + \o_{A}^2 b^\prime_{\pm} = 0 , \label{ideal-ode-b}
\end{align}
whose solutions are  
\begin{align}
v_{\pm} &= c_{1\pm} \cdot e^{ z_{1\pm} \cdot t} + c_{2\pm} \cdot e^{ z_{2\pm}  \cdot t} , \label{ideal-vtk}    \\
 b^\prime_{\pm}&=\frac{ i\o_{A} }{z_{1\pm}} c_{1\pm} \cdot e^{z_{1\pm} \cdot t}  + \frac{ i\o_{A} }{z_{2\pm}}  c_{2\pm} \cdot e^{z_{2\pm} \cdot t } \label{ideal-btk} ,\\
z_{1\pm} & \equiv \frac{i}{2}[ \pm\o_{I} + \sqrt{\o_{I}^2 + 4 \o_{A}^2  } ] , \qquad
z_{2\pm} \equiv \frac{i}{2}[\pm \o_{I}-\sqrt{\o_{I}^2+4 \o_{A}^2 } ] ,
\end{align}
where the coefficients $c_{1\pm}(\bk),c_{2\pm}(\bk)$ are assumed to be real functions of $\bk$, determined by initial conditions, and $b^\prime_{\pm}(t,\bk)$ is derived from \eq{spectrum-v} and \eq{ideal-vtk}. Since $\bv(t,\bx),\bb(t,\bx)$ are real functions, we have $\bv^*(t,\bk)=\bv(t,-\bk), \bb^*(t,\bk)=\bb(t,-\bk)$, which combined with \eqs{ideal-vtk}{ideal-btk}, imply $c_{1\pm}(-\bk)=c_{2\pm}(\bk)$. Then $|v_{\pm}|^2,|b_{\pm}|^2 ,v_{\pm}b^*_{\pm}$ can be expressed as  
\begin{align}
|v_{\pm}|^2 & = c_{1\pm}^2 + c_{2\pm}^2
+ 2(c_{1\pm} c_{2\pm}) \cdot \mathrm{cos} [\sqrt{\o_{I}^2 +4 \o_{A}^2} \cdot t]  \label{ideal-vtkvtk} , \\
|b^\prime_{\pm}|^2 & = ( \frac{ i\o_{A} }{z_{1\pm}})^2\cdot c_{1\pm}^2+ (\frac{ i\o_{A} }{z_{2\pm}})^2\cdot c_{2\pm}^2 - 2 (c_{1\pm}c_{2\pm}) \cdot \mathrm{cos} [\sqrt{\o_{I}^2+4\o_{A}^2 }\cdot t] 
 \label{ideal-btkbtk},\\
v_{\pm}b^{\prime *}_{\pm} & = \frac{ i \o_{A} }{z_{1\pm}} c_{1\pm}^2 + \frac{i\o_{A} }{z_{2\pm}} c_{2\pm}^2 - \frac{ (\pm\o_{I})\cdot c_{1\pm}c_{2\pm}}{\o_{A}} \mathrm{cos}[\sqrt{\o_{I}^2+4\o_{A}^2} \cdot t ]  - i \cdot \frac{ \sqrt{\o_{I}^2+4\o_{A}^2} \cdot c_{1\pm}c_{2\pm} }{\o_{A}} \mathrm{sin}[\sqrt{\o_{I}^2+4\o_{A}^2} \cdot t ]  
 \label{ideal-vtkbtk},
\end{align}
from which we can obtain the kinetic energy, magnetic energy, and various helicities, after integrating over $\bk$. The Eqs. (\ref{ideal-vtkvtk})-(\ref{ideal-vtkbtk}) exhibit the oscillatory behavior, even after the $\bk$ integration. Although $|v_{\pm}|^2$ or $|b_{\pm}|^2$ oscillate over time, $|v_{\pm}|^2+|b^\prime_{\pm} | ^2$ is constant, indicating energy conservation in ideal MHD. The last term in Eq. (\ref{ideal-vtkbtk}) is imaginary, but it vanishes after $\bk$ integrtion because $c_{1\pm}(-\bk)c_{2\pm}(-\bk)=c_{2\pm}(\bk)c_{1\pm}(\bk)$ and $\o_A(-\bk)=-\o_A(\bk)$, showing that this imaginary term has no contribution to cross helicity.

We now turn to \eq{2nd-ode-v}. It has the following characteristic equation
\begin{align}
\z^2 -(\pm i\o_{I}-\eta\bk^2)\cdot \z - [ i \eta \bk^2 (\o_{A} \xi_{\o}^{\prime} \pm \o_{I}) -\o_{A}^2 ] =0  \label{character-equation} ,
\end{align}
which has four solutions 
\begin{align}
\z = \z_{\pm}^{\a} & \equiv  \frac{1}{2}[ (\a i\o_{I}-\g_{\eta} ) \pm \sqrt{ -\o_{I}^2 + \g_{\eta}^2 + 2 \g_{\eta} \cdot (\a i\o_{I}) +4i \g_{\eta} \cdot \xi_{\o}^\prime\o_{A}  - 4\o_{A}^2 } ]  \qquad (\a=\pm 1)   \label{character-value}.
\end{align}
Then, one obtains
\begin{align}
   v_{+}  & = c_{1+}e^{\z_{+}^{+} \cdot t} +c_{2+} e^{\z_{-}^{+}\cdot t} = c_{1+}e^{-i\o_{+}^{-} \cdot t} +c_{2+} e^{-i\o_{-}^{-}\cdot t}   \label{cmvi-vtk+}   , \\
   v_{-}  & = c_{1-}e^{\z_{+}^{-}\cdot t}  +c_{2-} e^{\z_{-}^{-}\cdot t}= c_{1-}e^{-i\o_{+}^{+}\cdot t}  +c_{2-} e^{-i\o_{-}^{+}\cdot t}  \label{cmvi-vtk-} .
\end{align}
It is sufficient to assume $c_{1\pm}, c_{2\pm}$ to be real functions of $\bk$. Combing with Eq. (\ref{spectrum-v}), one obtains $b_{\pm}$:
\begin{align}
 b_{+} & = \frac{\sqrt{\r}}{\o_{A}}
 (-\o_{+}^{-}-\o_{I})c_{1+}e^{-i\o_{+}^{-} \cdot t}  + \frac{\sqrt{\r}}{\o_{A}} (-\o_{-}^{-}-\o_{I})c_{2+} e^{-i\o_{-}^{-}\cdot t}   \label{cmvi-btk+} ,\\
b_{-} & =\frac{\sqrt{\r}}{\o_{A}}
 (-\o_{+}^{+} +\o_{I} ) c_{1-}e^{-i\o_{+}^{+}\cdot t} +\frac{\sqrt{\r}}{\o_{A}}(-\o_{-}^{+} +\o_{I})c_{2-} e^{-i\o_{-}^{+}\cdot t}  \label{cmvi-btk-}.
\end{align}

With $v_\pm$ and $b'_\pm$ obtained, the kinetic and magnetic energies $\mathcal{E}_{b,v}$ as well as various helicities $\mathcal{H}_{b,c,v}$ can be constructed straightforwardly. In order to illustrate the behavior of $\mathcal{E}_{b,v},\mathcal{H}_{b,c,v}$ qualitatively, we ignore the terms suppressed by $(...)e^{-\g_{\eta} \cdot t}$ as  $e^{-\g_{\eta}\cdot t}\rightarrow 0 $ quickly for large times. We then have
\begin{align}
|v_{+}|^2 & \approx c_{1+}^{2} e^{ 2\mathrm{Im}(\o_{+}^{-}) \cdot t} + c_{2+}^{2} e^{ 2\mathrm{Im}(\o_{-}^{-}) \cdot t}    \label{} ,  \\
|v_{-}|^2  & \approx c_{1-}^{2} e^{2\mathrm{Im}(\o_{+}^+)\cdot t} + c_{2-}^{2} e^{2\mathrm{Im}(\o_{-}^{+}) \cdot t }  \label{},\\
 \o_{A}^2 \cdot |b^{\prime}_{+}|^2  &\approx 
  \{ [\mathrm{Re}(\o_{+}^{-}) +\o_{I} ] ^2 +  [\mathrm{Im}(\o_{+}^{-})]^2 \} \cdot c_{1+}^{2} e^{2\mathrm{Im}(\o_{+}^{-})\cdot t}  + \{ [ \mathrm{Re}(\o_{-}^{-})+\o_{I} ]^2 + [\mathrm{Im}(\o_{-}^{-})]^2 \} \cdot c_{2+}^{2} e^{2\mathrm{Im}(\o_{-}^{-})\cdot t }     \label{},\\
   \o_{A}^2 \cdot |b^{\prime}_{-}|^2 &\approx \{ [\mathrm{Re}(\o_{+}^{+}) -\o_{I}]^2 + [\mathrm{Im}(\o_{+}^{+})]^2  \} \cdot c_{1-}^{2} e^{2\mathrm{Im}(\o_{+}^{+})\cdot t} + \{ [\mathrm{Re}(\o_{-}^{+})-\o_{I} ]^2 + [\mathrm{Im}(\o_{-}^{+})]^2 \} \cdot c_{2-}^{2}  e^{2\mathrm{Im}(\o_{-}^{+})\cdot t }  \label{},\\
  \o_{A}\cdot v_{+}b_{+}^{\prime \ *} & \approx [ -\mathrm{Re}(\o_{+}^{-})+ i \cdot \mathrm{Im}(\o_{+}^{-}) -\o_{I} ]\cdot c_{1+}^{2} e^{2 \mathrm{Im}(\o_{+}^{-}) \cdot t }  + [-\mathrm{Re}(\o_{-}^{-})+ i \cdot \mathrm{Im}(\o_{-}^{-})-\o_{I}] \cdot c_{2+}^{2} e^{2 \mathrm{Im}(\o_{-}^{-})\cdot t }    \label{},\\
\o_{A}\cdot v_{-}b_{-}^{\prime \ *} & \approx [ -\mathrm{Re}(\o_{+}^{+})+ i \cdot \mathrm{Im}(\o_{+}^{+}) +\o_{I} ]\cdot c_{1-}^{2} e^{2 \mathrm{Im}(\o_{+}^{+}) \cdot t }  + [-\mathrm{Re}(\o_{-}^{+})+ i \cdot \mathrm{Im}(\o_{-}^{+})+ \o_{I}] \cdot c_{2-}^{2} e^{2 \mathrm{Im}(\o_{-}^{+})\cdot t }   \label{}.
\end{align} 
Therefore, for constant $\x_\o>0$, following the results in Sec.~\ref{sec:cve}, the modes $|v_{+}|^2,|b_{+}^\prime|^2$ will increase continuously with time even when $\xi^\prime_{\o} \leq 1$, while the modes $|v_{-}|^2,|b_{-}^ \prime|^2$ are unstable only when $\xi^\prime_{\o}>1$. However, once we allow $\x_\o$ to evolve in time according to the \eq{numerical-4}, the instability will eventually be tamed, as  analyzed in Sec.~\ref{sec:evolu}.

\end{widetext}

\nocite{*}

\bibliography{R-CMVI}

\end{document}